\documentclass[11pt]{article}
\usepackage[a4paper,margin=2.05cm]{geometry}
\usepackage[T1]{fontenc}
\usepackage{lmodern}
\usepackage{microtype}
\usepackage{amsmath,amssymb}
\usepackage{graphicx}
\usepackage{booktabs,tabularx,array}
\usepackage{authblk}
\usepackage{enumitem}
\usepackage{xcolor}
\usepackage[hidelinks]{hyperref}
\usepackage[authoryear,round]{natbib}
\let\parencite\citep
\let\textcite\citet
\hypersetup{pdftitle={Untangling EIT Waves: What a Measured Speed Actually Traces},pdfauthor={Olena Podladchikova},pdfsubject={Perspective for Solar Physics}}
\newcommand{\kms}{\mathrm{km\,s^{-1}}}
\newcommand{\Rs}{R_{\odot}}
\newcommand{\Efront}{E_{\rm front}}
\newcommand{\Erelease}{E_{\rm release}}

\newcolumntype{Y}{>{\raggedright\arraybackslash}X}

\title{Untangling EIT Waves:\\What a Measured Speed Actually Traces}
\author[1,2]{Olena Podladchikova}
\affil[1]{National Technical University of Ukraine ``Igor Sikorsky Kyiv Polytechnic Institute'', 37 Beresteiskyi Avenue, 03056 Kyiv, Ukraine}
\affil[2]{Leibniz Institute for Astrophysics Potsdam (AIP), An der Sternwarte 16, 14482 Potsdam, Germany}
\affil[]{Corresponding author: \href{mailto:epodlad@gmail.com}{epodlad@gmail.com}}
\date{}

\begin{document}
\maketitle

\begin{abstract}
Nearly three decades after the first SOHO/EIT observations, reported EUV-wave speeds still range from a few tens to more than one thousand kilometers per second. This Perspective asks first what those numbers actually trace. A measured image speed is produced by a chain that includes passband and line-of-sight weighting, cadence, difference imaging, the selected crest or leading edge, propagation sector, fitted interval, and projection geometry. Published same-event comparisons quantify the effect. For events common to the Nitta and Muhr analyses, early or fastest-sector measurements were typically higher: five of six directly shared events differed by about 30\%, while the larger 21-event overlap averaged $345\,\kms$ versus $296\,\kms$. For the 19 May 2007 disturbance, the reported 171~\AA{} peak speed was $475\pm47\,\kms$, compared with $238\pm20\,\kms$ in 304~\AA{} and a similar 195~\AA{} evolution; the 171~\AA{} cadence was four times faster than the 195~\AA{} cadence. A separate numerical audit of the published 1 April 2017 SWAP sampling shows that a repeated interval speed near $834\,\kms$ is almost exactly one 91.6-Mm radial ring per 110-s image interval. In a reconstructed sector of the 3 April event, an AIA 171~\AA{} ridge gives about $405\,\kms$, compared with the published SWAP sector mean of $484\,\kms$; an AIA 193~\AA{} ridge gives only about $250\,\kms$ while automatic exposure control strongly shortens the 193~\AA{} exposures. These are quantitative differences imposed on the measured observable, not automatic evidence for different MHD modes. A six-channel same-event pilot then gives a compact-source thermal excess of $1.5\times10^{29}$--$1.0\times10^{30}$ erg, a dimming/ejecta mechanical proxy of $4\times10^{27}$--$3\times10^{28}$ erg, and $6\times10^{25}$--$2\times10^{26}$ erg for the kinetic component in one fixed front-sector segment. The latter is a weak-compression, geometry-dependent estimate, not a total wave energy, and its small DEM enhancement is comparable to control-region variability. At smaller scales, published mini-wave speeds of about 14 and $45\,\kms$ yield illustrative weak-compression kinetic energies near $10^{22}$ and $10^{23}$--$10^{24}$ erg; these are model-dependent estimates, not measured thermal energies. The available measurements do not yet justify fitting a universal power law. They do define a falsifiable hypothesis: does $\Efront$ scale linearly with $\Erelease$, or does $\Efront/\Erelease$ change systematically from compact quiet-Sun events to global waves and shocks? Establishing that dependence would reveal whether eruption-to-front coupling is scale invariant and, together with the event-energy distribution and the fraction dissipated in the corona, which eruption scales can contribute materially to coronal heating.
\end{abstract}

\section{Why another synthesis after nearly three decades?}
The Extreme-ultraviolet Imaging Telescope (EIT) was designed to image the solar atmosphere in several temperature-sensitive EUV bands and opened a global view of the low corona \parencite{Delaboudiniere1995}. The 12 May 1997 Earth-directed eruption then became a defining example: EIT recorded source-region dimming, post-eruption arcades, and a bright front propagating across the disk, while LASCO later observed the associated halo CME \parencite{Thompson1998}. Since then, the subject has been reviewed repeatedly and in depth \parencite{ZhukovAuchere2004,PatsourakosVourlidas2012,LiuOfman2014,Warmuth2015,Long2017}. The need today is therefore not another encyclopedic inventory of every proposed ``EIT-wave model.''

A narrower problem remains unresolved in practice: \emph{what exactly does a published EUV-wave speed measure?} The underlying MHD-wave framework is well established: Alfv\'en, slow and fast magnetoacoustic responses have distinct propagation, polarization and dispersive properties \parencite{NakariakovVerwichte2005}; the unresolved step here is identifying which physical response, if any, a particular image front traces. Different catalogs use different cadences, passbands, sectors, front definitions and fit intervals. High-cadence observations further show that one eruption can contain several moving ridges \parencite{Liu2010,Liu2012}. A number such as $200$, $500$, or $1000\,\kms$ is consequently not, by itself, an MHD-mode label.

Catalogues, curated statistical samples, automated detection pipelines, and model-assisted shock-characterization products are also distinct observational products. Agreement or disagreement among them cannot be judged from one event count or one speed column unless their selection rules, geometry, tracked feature, and derived quantities are first matched.

This distinction also matters in practice. SDO provides high-cadence full-disk EUV imaging but carries no coronagraph of its own. An EUV wave or dimming can therefore supply low-coronal evidence of an eruption before, or independently of, clear coronagraph confirmation. A recent space-weather synthesis identifies a large-scale EUV wave as strong evidence that a CME has occurred and notes that wave direction or rotation may contain useful forecasting information, although statistical validation is still required \parencite{Green2026}. The same ambiguity that complicates wave physics therefore also affects CME interpretation.

\begin{figure}[t]
\centering
\includegraphics[width=0.99\linewidth]{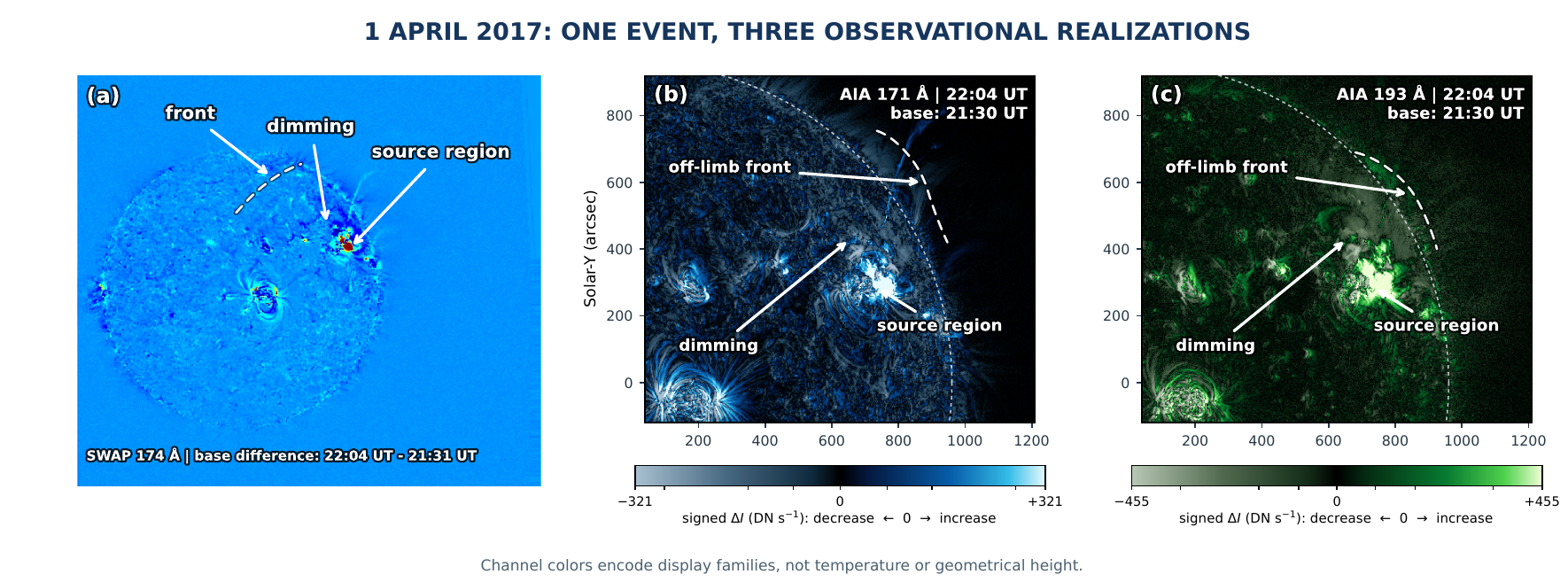}
\caption{One event, three signed base differences on 1 April 2017. (a) SWAP 174~\AA{} at 22:04~UT relative to 21:31~UT, independently constructed for this Perspective from the event analyzed by \textcite{OHara2019}. It is used as a qualitative display raster, so no quantitative SWAP color bar is given. (b,c) Exposure-normalized AIA 171 and 193~\AA{} differences at approximately 22:04~UT relative to 21:30~UT, with $\Delta I=I_{\rm event}/t_{\rm exp,event}-I_{\rm base}/t_{\rm exp,base}$ in DN~s$^{-1}$. Negative, zero, and positive values indicate dimming, little change, and brightening. Channel-specific symmetric clipping and signed asinh scaling are used; no ordinary logarithm is applied. Arrows mark the source region, dimming, and front. Dashed curves are qualitative guides, not fitted masks: SWAP marks the visible on-disk segment and AIA the off-limb front; the guides correspond qualitatively to the locations analyzed by \textcite{OHara2019}. Differences among panels show how channel response and processing alter the visible crest or leading edge. Blue denotes the 171/174~\AA{} family and green the 193/195~\AA{} family only; the colors are not a temperature scale. No cross-instrument speed is derived.}
\label{fig:sequence}
\end{figure}

The observational problem can be seen directly in one event (Fig.~\ref{fig:sequence}). The SWAP difference emphasizes the on-disk source--dimming--front system, whereas AIA makes the off-limb front much clearer; AIA 171 and 193~\AA{} also differ in front contrast and morphology. These differences do not by themselves imply different physical waves: channel response, instrumental sampling, and image processing determine which part of the evolving disturbance becomes visible. Detailed feature identification and kinematic analysis of this event are presented by \textcite{OHara2019}; Fig.~\ref{fig:sequence} is used here only to show this observation-operator problem, and no new cross-instrument speed is derived. A speed comparison becomes meaningful only after the physical structure and selected crest or leading edge have been matched.

At smaller scales, published mini-eruptions share the observable anatomy of a compact source, ejecta or dimming, and an outward front \parencite{Innes2009,Podladchikova2010}. That morphological continuity does not prove a common MHD mode or a constant transported fraction. It identifies the missing measurement: how the released energy is divided among compact heating, ejecta, and the propagating front as eruption scale decreases, and therefore whether eruption-to-front coupling is scale invariant.

The working thesis is simple. Before a physical label is assigned, one must separate the structure being tracked, the MHD response, the driver history, and the observation operator. Topology and energy partition then become constraints rather than additional names for the same speed. Untangling the speed is required before a universal cross-scale energy relation can be tested.

\section{A measured EUV speed is an output, not a primitive}
For an optically thin EUV passband, the recorded intensity can be written schematically as
\begin{equation}
 I_\lambda(x,y,t)\propto\int n_e^2\,G_\lambda(T,n_e)\,dl,
 \label{eq:intensity}
\end{equation}
where $G_\lambda$ is the instrumental temperature response. The image is therefore not a density map. A propagating compression can change density and temperature simultaneously; a dimming can include evacuation, temperature migration out of a passband, or both; and line-of-sight superposition can mix structures at different heights.

\begin{figure}[t]
\centering
\includegraphics[width=0.95\linewidth]{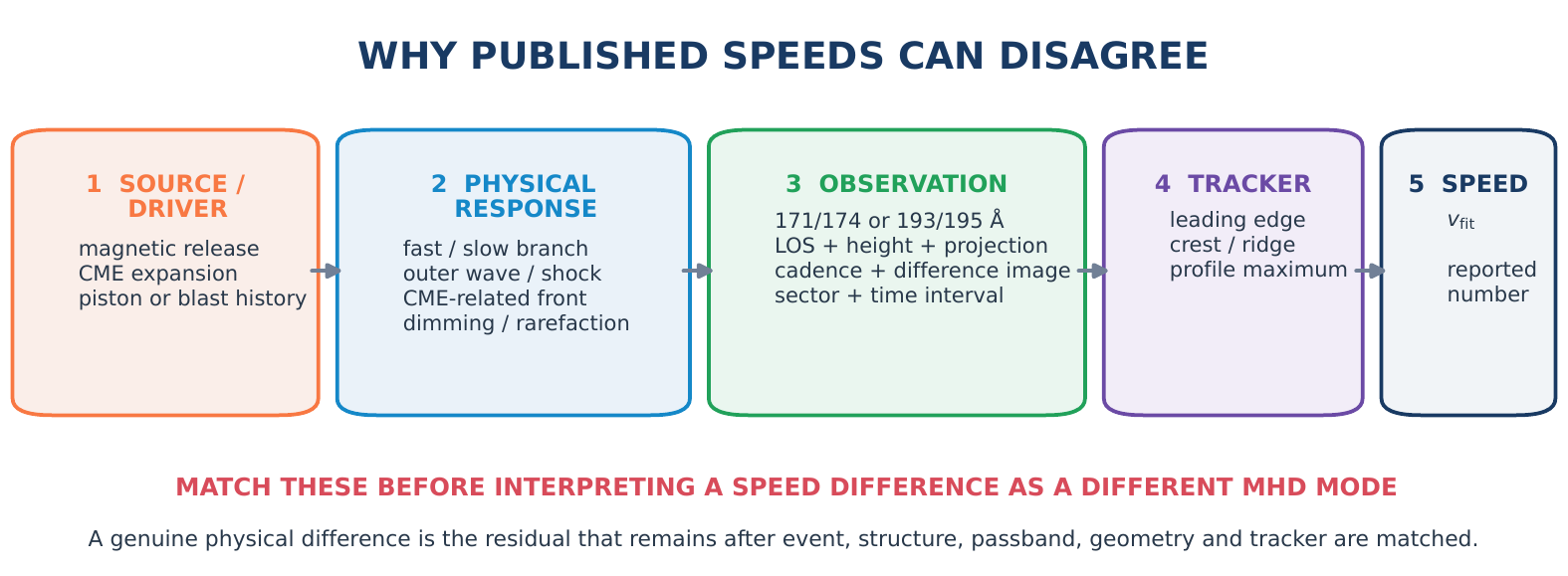}
\caption{Causal chain from eruption to a published speed. Driver history, MHD response, CME-related restructuring, passband and line-of-sight weighting, and the selected tracker are separate levels. A speed difference can be interpreted as physical only after event, structure, geometry, channel, and tracker are matched; a passband itself has no Alfv\'en speed.}
\label{fig:operator}
\end{figure}

The measurement chain is shown in Fig.~\ref{fig:operator}. Magnetic release and CME expansion supply the driver; ``piston'' and ``blast'' describe the driver history, not the MHD branch. The plasma response may then contain a fast or slow branch, a shock, a CME-related front, dimming or rarefaction. Only after that physical response is projected through a temperature-sensitive passband and the line of sight do WCS co-registration, difference imaging, the selected crest or leading-edge tracker, and the final fit transform an intensity pattern into a reported speed. In this review, ``ridge'' is reserved for the line that the selected feature produces in a time--distance diagram. Two analyses can therefore begin with the same eruption and still measure different, internally valid observables.

There is no ``Alfv\'en speed of 171~\AA'' or ``Alfv\'en speed of 195~\AA.'' The local Alfv\'en speed, $v_A=B/\sqrt{\mu_0\rho}$, belongs to the plasma. Different passbands can nevertheless emphasize different parts of the same disturbance---for example plasma at different temperatures, heights, densities or magnetic geometries---and thereby select features propagating through regions with different $v_A$ or fast-mode speed. The distinction is between \emph{what a channel makes visible} and \emph{the underlying plasma physics}.

\section{Front and dimming should be analyzed as one structured event}
\begin{figure}[t]
\centering
\includegraphics[width=0.88\linewidth]{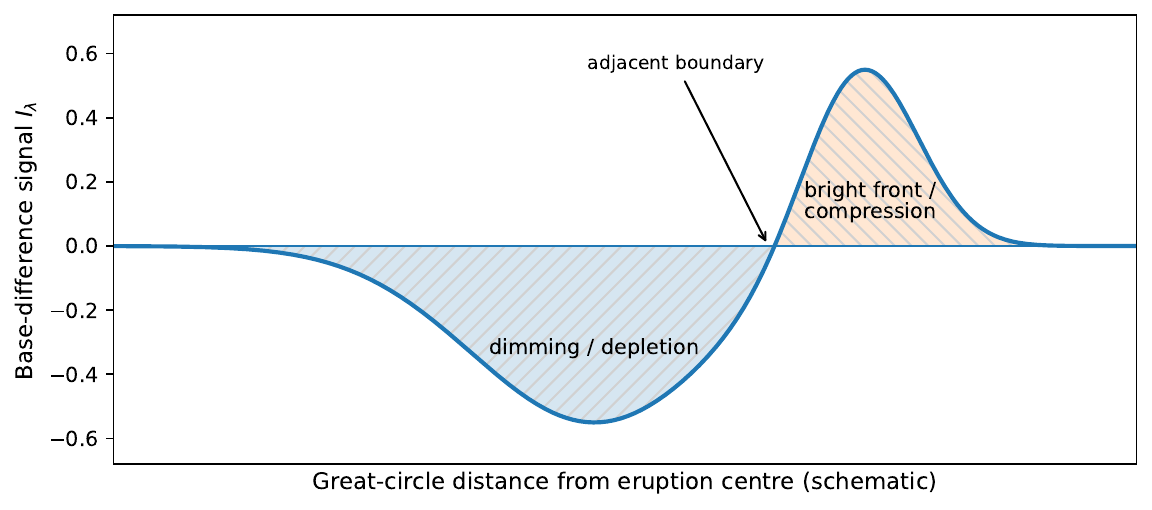}
\caption{Idealized front--dimming coupling: an outward positive emissivity perturbation followed by a trailing negative perturbation. This is the simplest quasi-isotropic case. NEMO also distinguished dipolar, quadrupolar, and irregular dimmings; these are topology-sensitive morphologies, not evidence that real events are radially symmetric.}
\label{fig:nemo}
\end{figure}

The early Novel EIT Wave Machine Observing (NEMO) work \footnote{NEMO operated at the Royal Observatory of Belgium as a real-time alert pipeline using the approximately 12-min-cadence SOHO/EIT 195~\AA{} CME-watch stream from 2006 until the alerts ended on 1 August 2010. Its historical website and EIT-wave and eruptive-dimming archive remain available at \url{https://www.sidc.be/nemo/}.}
treated the propagating bright front and the trailing dimming together rather than as unrelated products \parencite{PodladchikovaBerghmans2005,PodladchikovaBerghmans2005Energetic,Podladchikova2012}. In the 12 May 1997 event, the outer dimming boundary was found to adjoin the inner boundary of the bright front, and the integrated positive and negative difference-image signals evolved with an approximate temporal balance. That result still constrains interpretation, but its modern meaning must be precise: it is a coupling of \emph{EUV emissivity perturbations}, not a proof of mass or energy conservation, because of Eq.~\eqref{eq:intensity}. Fig.~\ref{fig:nemo} summarizes this coupled morphology schematically.

Modern DEM studies strengthen the physical side of the dimming picture. Large coronal dimmings are dominated by genuine density depletion and plasma evacuation, although temperature changes contribute to the observed intensity \parencite{Vanninathan2018,Dissauer2018}. Thus, a successful eruption model should not reproduce only the position of a bright ridge; it should also explain the spatial and temporal relation between compression, depletion and the evolving CME footprint.

Rotation and asymmetry can reveal topology. Attrill and collaborators found opposite senses of apparent coronal-wave rotation for source regions of opposite helicity and later showed that the rotation sense was consistent with the source-region helicity in events observed by EIT, EUVI and AIA \parencite{Attrill2007,Attrill2014}. This relation constrains interpretation but is not a unique mode discriminator. Rotation may trace CME-footprint migration, reconnection and magnetic topology, while a detached MHD front may simultaneously propagate through the structured fast-mode-speed landscape. The two components need not be mutually exclusive \parencite{Cohen2009,Cohen2010}.

The same morphology also extends to smaller eruptions. The STEREO mini-eruption analyzed by \textcite{Podladchikova2010} produced a faint front and diffuse dimming with a spatial scale far below classical global EIT events; the selected inner front boundary propagated on average at about $14\,\kms$. Independently, \textcite{Innes2009} estimated about 1400 quiet-Sun mini-CME-like events per day over the whole Sun, roughly one third with faint propagating disturbances. Similarity across scale is evidence for a common organizing role of eruptive magnetic restructuring, but it does not imply that the same MHD mode or the same energy fraction is present at every scale.

\section{The quantitative speed result: why catalogs and wavelengths disagree}
\begin{figure}[t]
\centering
\includegraphics[width=0.99\linewidth]{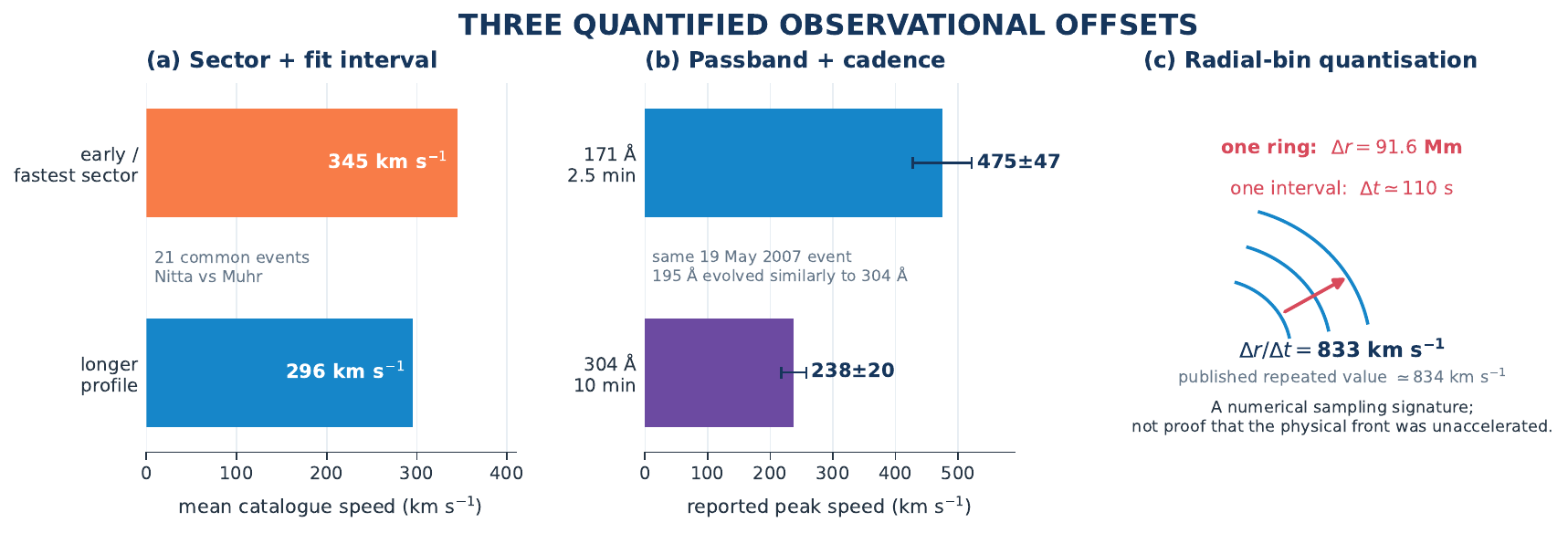}
\caption{Three measurement-dependent speed offsets. (a) Across 21 common events, the Nitta early or fastest-sector speeds averaged about $345\,\kms$, versus $296\,\kms$ from the longer Muhr profiles; these are sample means, not two directions in one event. (b) For 19 May 2007, 171~\AA{} gave $475\pm47\,\kms$ at 2.5-min cadence, while 304~\AA{} gave $238\pm20\,\kms$ and 195~\AA{} was similar at about 10-min cadence. (c) For 1 April 2017, one 91.6-Mm SWAP ring per approximately 110~s gives $833\,\kms$, matching the repeated $834\,\kms$ interval speed.}
\label{fig:sampling}
\end{figure}

This Perspective combines published same-event and common-event comparisons with a numerical audit of the published SWAP sampling and one controlled AIA sector reconstruction. The main result is that changing the fitted interval, propagation sector, cadence, passband response, exposure control, or selected front can change the reported speed by tens of percent and, in some cases, by about a factor of two. Fig.~\ref{fig:sampling} collects three quantified examples. The AIA reconstruction is used as an observation-operator test, not as a definitive cross-instrument recalibration of the published SWAP result.

Warmuth and Mann used speed and acceleration to identify three kinematic populations: initially fast/decelerating (class~1), intermediate/nearly constant (class~2), and very slow/erratic (class~3) events \parencite{WarmuthMann2011,Warmuth2015}. They interpreted classes~1 and 2 as nonlinear fast-mode waves or shocks and linear fast-mode waves, respectively, whereas class~3 could reflect magnetic reconfiguration or, as a hypothesis, a slow-mode response. Nitta et al., by contrast, visually selected AIA 193~\AA{} large-scale propagating fronts, measured them in several directions during the early propagation, and retained the \emph{highest} directional speed for each event \parencite{Nitta2013}. They found no evidence for a multi-class population or for a systematic tendency of faster fronts to decelerate and slower fronts to accelerate. The two studies therefore did not recover the same kinematic partition, and their reported speeds differ because their measurement definitions differ.

The Muhr sample adds a third result. For selected strong, well-pronounced waves, \textcite{Muhr2014} found a smooth transition near a start speed of $230\,\kms$ between slower, nearly constant propagation and faster, decelerating propagation, and interpreted the sample as fast-mode MHD waves. Weak events comparable to Warmuth's class~3 were excluded by the selection. Muhr et al. also compared common events with the Nitta analyses. For five of six events shared with the 2013 AIA sample, the Nitta speed was about 30\% higher. Across 21 events common to the Nitta 2013/2014 studies and the Muhr sample, the mean difference was $56\pm53\,\kms$; the Nitta values averaged about $345\,\kms$, compared with $296\,\kms$ in the Muhr analysis. Muhr et al. identified two causes: the Nitta fits emphasized the first $15$--$20$ min, before later deceleration, and the analyses sometimes used different propagation sectors; the 2013 catalog explicitly selected the fastest sector. The discrepancy therefore reflects the measurement definitions and does not necessarily imply different plasma physics. Table~\ref{tab:catalogs} summarizes these three measurement definitions and their immediate consequences.

\begin{table}[t]
\centering
\caption{Three ways of turning an EUV disturbance into a speed. The numerical differences matter because the definitions differ; a kinematic class or transition is not, by itself, an MHD-mode identification.}
\label{tab:catalogs}
\small
\begin{tabularx}{\linewidth}{@{}>{\raggedright\arraybackslash}p{2.6cm}>{\raggedright\arraybackslash}p{3.1cm}Y Y@{}}
\toprule
Study & Data / selection & What is measured & Consequence \\
\midrule
Warmuth--Mann & EIT/EUVI and related wave samples & Initial speed plus acceleration; three kinematic classes & Classes~1--2 were interpreted as nonlinear/linear fast-mode responses; class~3 remained physically ambiguous. \\
Nitta et al. 2013 & AIA 193~\AA{}, 138 on-disk LCPFs & Early propagation in several directions; highest speed retained & High-speed distribution from a fastest-direction measure; no multi-class population was found. \\
Muhr et al. 2014 & 60 strong STEREO/EUVI waves & Perturbation-profile tracking over a longer evolution & Smooth transition near $230\,\kms$; fast-mode interpretation for the selected strong sample; weak class~3-like events excluded. \\
\bottomrule
\end{tabularx}
\end{table}

A second demonstration uses one event observed in several passbands. For the 19 May 2007 STEREO/EUVI disturbance, \textcite{Long2008} obtained a peak speed of $475\pm47\,\kms$ in 171~\AA{}, whereas the 304~\AA{} result was $238\pm20\,\kms$ and the 195~\AA{} kinematics were comparable to the latter. The cadences differed: 171~\AA{} had a 2.5-min cadence, while 195~\AA{} had about 10 min. The higher cadence changed the inferred velocity by roughly a factor of two and the acceleration by roughly an order of magnitude. This same-event discrepancy shows the effect of the observation operator and should not immediately be interpreted as two MHD modes.

\subsection{A concrete sampling test: the 2017 SWAP events}
The extended-field SWAP observations of 1 and 3 April 2017 provide a clear sampling test \parencite{OHara2019}. Published mean speeds were about $417\,\kms$ on 1 April and $457$ and $484\,\kms$ in two sectors on 3 April. On 1 April, however, three consecutive interval speeds clustered near $834\,\kms$. The profiles used radial rings 91.6~Mm wide, while consecutive frames were separated by about 110~s. Dividing the 91.6-Mm ring width by the approximately 110-s image interval gives $833\,\kms$: one radial ring per image interval.
This does not prove that the front had no physical acceleration. It does show that the repeated short-interval value was strongly quantized by the spatial and temporal sampling.

As a limited follow-up observation-operator test, a matched-cadence AIA data set was prepared using uniform 2-min 171 and 193~\AA{} sequences: 19 frames per channel on 1 April and 20 per channel on 3 April. The purpose is not to remeasure the published SWAP kinematics, but to test whether the same front can be recovered in AIA under a common cadence and sector definition. Fig.~\ref{fig:sequence} shows one same-time AIA pair from that set beside the independently constructed SWAP base-difference display. The FITS World Coordinate System (WCS) maps image pixels to physical solar coordinates and is essential for this cross-instrument comparison because SWAP and AIA have different plate scales, reference pixels, pointing, and roll. In the reconstructed lower sector of the 3 April event, a central AIA 171~\AA{} ridge fit gives about $405\,\kms$, versus the SWAP sector mean near $484\,\kms$ reported by \textcite{OHara2019}. Allowing AIA 193~\AA{} to select its own strongest ridge gives about $250\,\kms$, but the 193~\AA{} automatic exposure control shortened exposures to approximately 0.02--0.37~s during the rapid evolution, strongly degrading the faint-front observable. The lower 193~\AA{} value therefore cannot be interpreted as a second MHD-mode speed. A safer protocol is to use SWAP 174 and AIA 171~\AA{} to identify one common cool-front trajectory, then sample AIA 193~\AA{} along that same trajectory. Because exact historical SWAP Level-1 files and original masks are not yet available, the offset between $405$ and $484\,\kms$ cannot be assigned to a single cause: WCS and sector reconstruction, temporal sampling, and crest selection can all contribute. It therefore shows how measurement choices can shift the inferred speed, but it is not an exact AIA--SWAP comparison. At this stage, 1-min AIA data would not remove these dominant uncertainties; exact WCS transfer, exposure history, and crest identity matter more than doubling an already 2-min cadence.

For a different event, the 15 February 2011 wave, \textcite{Vanninathan2015} used DEM analysis to show that plasma visible in AIA 171~\AA{} was heated into the temperature range emphasized by AIA 193 and 211~\AA{}, while density increased by about 6--9\% and temperature by about 5--6\% at the front. Thus, the same physical compression may brighten in 193/211~\AA{} while weakening or darkening in 171~\AA{}: the channel selects the visible thermodynamic response, not the governing MHD physics.

\subsection{Radial propagation and azimuthal pattern drift are separate observables}
An additional same-event test follows from the apparent rotation first quantified in the NEMO analysis. For the 12 May 1997 event, \textcite{PodladchikovaBerghmans2005} reported three radial speeds of 258, 225 and $258\,\kms$. Tracking the weighted centers of localized running-difference intensity regions gave angular rates of $2.06$, $5.48$, $2.64$ and $2.26\times10^{-4}\,\mathrm{rad\,s^{-1}}$ for the available southeast and northwest components. The corresponding resultant pattern speeds were 263, 365, 265 and $313\,\kms$. These are velocities of intensity-localization centers on the reconstructed solar surface, not Doppler measurements of plasma flow.

\begin{figure}[t]
\centering
\includegraphics[width=0.99\linewidth]{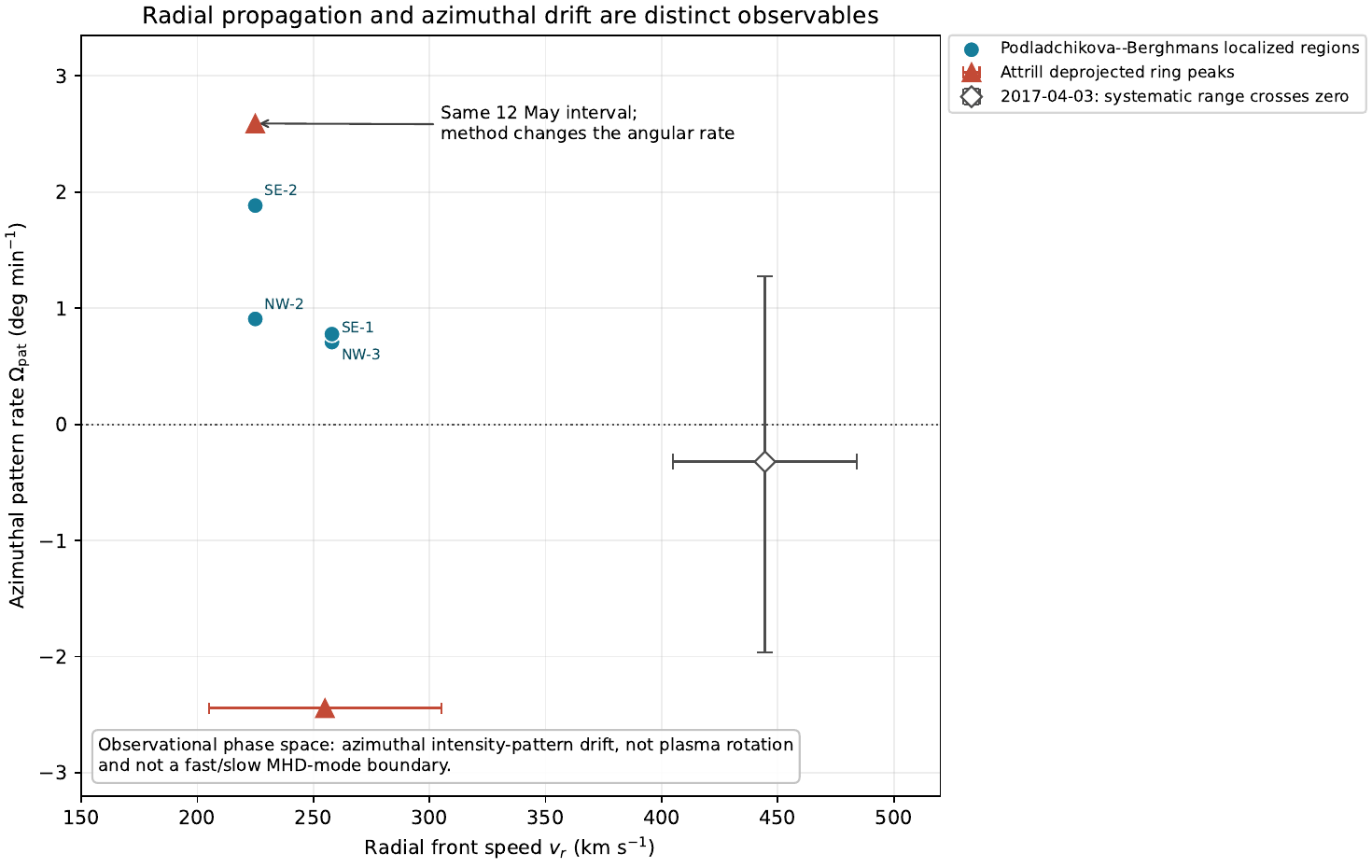}
\caption{Observational radial--azimuthal plane. Filled circles show the localized intensity-region measurements of \textcite{PodladchikovaBerghmans2005}; triangles combine the deprojected ring-peak shifts of \textcite{Attrill2007} with matched or published radial speeds; the open diamond is the 3 April 2017 sensitivity result, whose systematic interval crosses zero. The ordinate is signed drift of an EUV intensity pattern, not plasma rotation. This is not a fast/slow-mode map; it is a compact set of observables against which candidate models can be tested.}
\label{fig:radaz}
\end{figure}

\textcite{Attrill2007} used a different observable: the phase shift of peaks in a deprojected base-difference ring-intensity profile. The 12 May peak shifted $44^\circ$ counterclockwise in 17 min, while two peaks in the 7 April event shifted $22^\circ$ clockwise in 9 min. Pairing the latter with the published event-average radial speed $255\pm50\,\kms$ \parencite{Thompson1999} gives the two opposite-sign points in Fig.~\ref{fig:radaz}. During the same 05:07--05:24~UT interval on 12 May, the localized-region method gives about 0.91 and $1.88^\circ\,\mathrm{min^{-1}}$, whereas the deprojected ring peak gives $2.59^\circ\,\mathrm{min^{-1}}$. The difference is a feature-definition result within one event.

The same procedure was tested on the reconstructed lower sector of the 3 April 2017 event. A centroid measured within an outward-moving annular region in AIA 171~\AA{} gave a preferred rate of $+1.12\pm0.78^\circ\,\mathrm{min^{-1}}$, but an 81-member sensitivity ensemble spanning the allowed ridge position, radial speed and ring width gave a median of $-0.32^\circ\,\mathrm{min^{-1}}$ and a 16--84\% range of $-1.96$ to $+1.27^\circ\,\mathrm{min^{-1}}$. Because the sign is not stable, this is a non-detection of systematic azimuthal drift, not a measured rotation.

Classical wave theory shows why several observables are useful. Whitham's construction for small-amplitude disturbances in ideal MHD displays distinct fast and slow magnetosonic wavefronts, whose shapes depend on propagation direction relative to the background magnetic field \parencite[Sec.~7.9, pp.~259--262, Figs.~7.14--7.15]{Whitham1974}. The planar piston and arbitrary-discontinuity problems of \textcite[Chap.~V, Secs.~1--2]{KulikovskiyLyubimov1965} go further: for fixed thermodynamic and magnetic states, maps in piston-velocity or initial velocity-jump space organize the admissible combinations of fast and slow shocks, fast and slow rarefaction waves, rotational discontinuities, contact discontinuities and, after sufficiently strong rarefaction, vacuum formation. These are theoretical state-space diagrams, not diagrams of remotely measured front speeds. They show that several independent observables can restrict the admissible solution family more tightly than one speed can.

Within the broader Sedov--Taylor point-blast tradition \parencite{Sedov1959}, Korobeinikov's framework for conducting, magnetized gas explicitly distinguishes the fast and slow magnetosonic characteristic branches and classifies the corresponding fast and slow MHD shocks through their jump and evolutionary conditions \parencite{KorobeinikovKarlikov1960}; see also \textcite[Chap.~I, Secs.~3.3 and 4.1; Chap.~VII, Secs.~4--5]{Korobeinikov1973,Korobeinikov1991}. In special magnetic configurations, point-blast problems admit self-similar reductions, whereas a constant background field introduces anisotropy and can require perturbative or numerical treatment. These idealized impulsive problems provide analytical limiting cases and benchmarks for MHD calculations, not a model of the full flare--CME driver history. The framework therefore makes a fast/slow two-front interpretation physically admissible and falsifiable, but it does not predict that every point blast produces two distinct shocks. Genuinely co-temporal, spatially distinct fronts propagating differently may be treated as candidates for separate fast and slow responses; they become candidates for identification as a fast/slow shock pair only if both satisfy compression and characteristic-speed diagnostics, including the propagation direction relative to the magnetic field and the density, pressure and magnetic-field jumps.

This theory suggests a clear remote-sensing experiment. The coordinates in Fig.~\ref{fig:radaz} are the radial speed of a selected image feature and the signed azimuthal drift of an intensity pattern. They are neither plasma-flow components nor the initial velocity jumps used by \textcite[Chap.~V, Sec.~2]{KulikovskiyLyubimov1965}, and they do not identify an MHD branch by themselves. If both are measured robustly for the same physical structure and combined with compression and density-depletion diagnostics, three-dimensional front geometry, an upstream magnetic-field estimate and front--driver separation, candidate MHD solutions can be passed through the same observation operator and compared with the data. Models that cannot reproduce the joint radial and azimuthal behavior can then be rejected. Fig.~\ref{fig:radaz} is therefore a target for model discrimination, while fast/slow identification remains conditional on characteristic-speed, magnetic-geometry and jump diagnostics.

The practical rule is therefore: when two reported speeds differ, first match the event, sector, time interval, ridge definition, cadence and geometry. Only the residual should be treated as physical.

\section{Mode family, driver history, and several fronts}
\label{sec:modes}
\begin{figure}[t]
\centering
\includegraphics[width=0.90\linewidth]{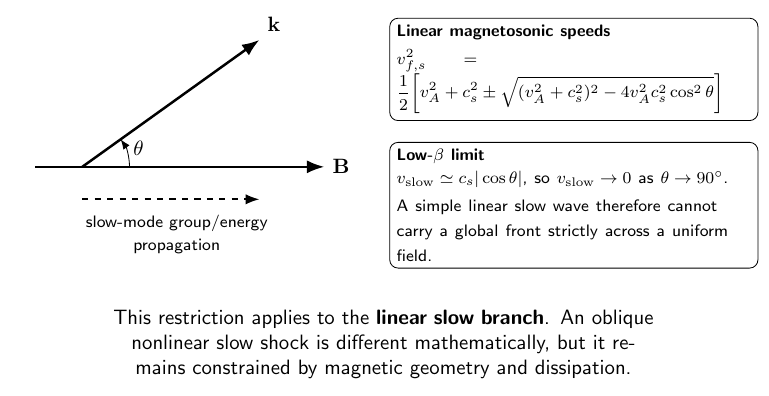}
\caption{Angular constraint on the slow magnetosonic branch. Slow waves are physically real and well observed in field-guided coronal structures, but low projected speed alone cannot identify a global slow mode. A global interpretation must also be compatible with magnetic geometry and thermodynamic or jump diagnostics.}
\label{fig:slowgeometry}
\end{figure}

The observation operator explains why measured speeds can differ, but it does not remove the need for MHD interpretation. In a low-$\beta$ corona, the slow magnetosonic speed is approximately $c_s|\cos\theta|$, where $\theta$ is the angle between propagation and the magnetic field; a nearly cross-field fast disturbance instead propagates near the local fast-mode speed, approximately $(v_A^2+c_s^2)^{1/2}$. A low image-plane speed is therefore not proof of the slow branch. A slow mode reaches a speed of order $c_s$ only when propagation is sufficiently field-aligned, whereas a quasi-isotropic global front crosses a wide range of field directions. Conversely, $200$--$300\,\kms$ does not exclude a fast disturbance in a structured quiet corona, especially when a finite-amplitude front has weakened or low cadence misses its early evolution \parencite{WillsDavey2007,Warmuth2015}. Fig.~\ref{fig:slowgeometry} makes this angular constraint explicit.

Here $\theta$ describes the wave-propagation direction relative to the local magnetic field. Shock obliquity is conventionally written as $\theta_{Bn}=\cos^{-1}(|\mathbf{B}_1\mathbin{\cdot}\hat{\mathbf{n}}|/|\mathbf{B}_1|)$, where $\hat{\mathbf{n}}$ is the local shock normal and $\mathbf{B}_1$ is the upstream magnetic field. The conventional division calls a shock quasi-parallel when $\theta_{Bn}<45^\circ$ and quasi-perpendicular when $\theta_{Bn}>45^\circ$. These terms describe magnetic geometry; they are not synonyms for the fast and slow shock families. In situ measurements can constrain the local field, shock normal, and plasma jumps more directly. Applying the same classification to a remote coronal front requires a three-dimensional front normal and an upstream coronal-field estimate; an image-plane propagation direction alone is insufficient. Thus, piston/blast, fast/slow, and quasi-parallel/quasi-perpendicular answer three different questions: driver history, MHD family, and shock obliquity. Compression and Mach-number diagnostics then determine whether a shock interpretation is supported.

The historical slow-shock--plus--rarefaction interpretation of selected NEMO front--dimming pairs remains useful when stated as a falsifiable hypothesis. One impulsive initial-value problem can yield a compressive structure followed by an under-pressure or depleted structure. A nonlinear slow shock, however, must satisfy the magnetic geometry, Rankine--Hugoniot jumps and observed propagation direction; a proposed rarefaction must be separated from CME evacuation using timing, DEM or spectroscopy, and front--CME geometry. Low speed, rotation, or dimming alone is not a mode discriminator. Modern piston/CME models can also generate source-region and trailing rarefactions \parencite{Lulic2013}.

The classical Petschek reconnection model provides an independent theoretical context for slow shocks: standing slow-mode shocks bound the reconnection exhaust \parencite{Petschek1964}. Together with the Korobeinikov framework \parencite[Chap.~I, Sec.~4.1]{Korobeinikov1973,Korobeinikov1991}, this establishes slow shocks as physically admissible MHD structures in coronal models, but it does not identify a global propagating EUV front as a slow shock from image-plane speed alone.

The words ``piston'' and ``blast'' describe driver history, not MHD branch. Rapid lateral CME expansion may first push and steepen a compression. As the driver weakens, the outer front can detach and propagate more freely while decelerating \parencite{Veronig2010,Lulic2013}. Observational blast-wave treatments have also combined a strong-shock self-similar approximation with EUV/Moreton and type-II kinematics, while nonlinear geometrical acoustics provides the complementary weak-shock limit \parencite{Grechnev2011,AfanasyevUralov2011}. The appropriate observable is therefore the evolving separation and amplitude of the front relative to the CME flank, rather than the label alone.

Finally, ``one eruption = one front = one speed'' is not generally valid. Erupting-flux-rope calculations produce an outer fast wave or shock together with a slower CME-related front; global simulations add restructuring and stationary components; and high-cadence AIA observations resolve multiple sharp and diffuse fronts \parencite{Chen2002,Chen2005,Cohen2009,Liu2010,Liu2012}. Events within this observational class that show two genuinely co-temporal, spatially distinct compressive fronts are therefore candidate data sets for the Korobeinikov-motivated fast/slow test \parencite[Chap.~I, Secs.~3.3 and 4.1]{Korobeinikov1973,Korobeinikov1991}, but two image-plane speeds alone are insufficient: repeated ridge selection, cadence aliases, passband-dependent crest shifts, and the fast-wave-plus-CME-front alternative must first be excluded. The order of inference is therefore: identify the crest or leading edge, reconstruct its measurement operator and relation to the driver, and only then use speed, compression and geometry to classify the MHD response.

\section{Diagnostic criteria for identifying a shock: speed alone is insufficient}
The shock criterion can be stated directly. A bright EUV front should be called a shock only when a density compression $X=n_2/n_1>1$ is demonstrated and the inferred disturbance is super-magnetosonic (or, under appropriate low-$\beta$ assumptions, super-Alfv\'enic). Under the low-$\beta$, perpendicular-shock approximation with $\gamma=5/3$, the Rankine--Hugoniot relation may be written \parencite{Vrsnak2002}
\begin{equation}
 M_{A,\perp}=\left[\frac{X(X+5)}{2(4-X)}\right]^{1/2}.
 \label{eq:shockcriterion}
\end{equation}
The assumptions in Eq.~\eqref{eq:shockcriterion} matter: it does not apply unchanged at arbitrary plasma $\beta$ or $\theta_{Bn}$. If $X$ is estimated from EUV intensity alone via $X\simeq\sqrt{I_2/I_1}$, the approximation requires an approximately unchanged temperature response and line-of-sight depth. DEM-derived density is preferable when the front produces measurable heating.

A literature event illustrates this criterion. \textcite{Long2015Energy} combined EUV and radio observations of the 25 February 2014 eruption, obtained a Mach number greater than unity, and then estimated an initial shock-wave energy of about $2.8\times10^{31}$ erg, roughly 10\% of the associated CME kinetic energy. Here, ``shock'' was supported by an independent Mach-number diagnostic rather than inferred from a large image-plane speed.

The same discipline applies at the slow end. A speed of $50$ or $150\,\kms$ does not establish the slow MHD branch. Slow-mode identification additionally requires propagation relative to the magnetic field, thermodynamic phase relations and a characteristic speed consistent with the local plasma. The historical slow-shock plus rarefaction interpretation of some EIT front--dimming pairs is therefore retained as a testable hypothesis, not as a conclusion from velocity alone.

\section{Energetics across scales: from estimates to a falsifiable scaling law}
The energy question becomes unclear when the compact source is mixed with its propagating response. A magnetic eruption partitions released free energy schematically as
\begin{equation}
 \Erelease=E_{\rm heat}+E_{\rm ejecta}+\Efront+E_{\rm nonth}+\cdots,
 \label{eq:partition}
\end{equation}
where $E_{\rm nonth}$ denotes the energy carried by accelerated nonthermal particles. Dimming is an observational proxy for part of the evacuation/ejecta evolution, not an extra conserved energy reservoir. The terms in Eq.~\eqref{eq:partition} are physically different. In particular, a measured source-region thermal energy is neither a measurement of $\Erelease$ nor of $\Efront$.

At the small-scale end, mini-eruptions provide a poorly quantified extension of the EUV-wave phenomenon. \textcite{Podladchikova2010} reported a quiet-Sun front propagating over about 40~Mm in 20~min with a mean apparent speed near $14\,\kms$, while \textcite{Innes2009} found faint fronts with typical speeds near $45\,\kms$, lifetimes of about 30~min, and travel distances near 80~Mm. From their estimated 1400 mini-eruptions per day and one-third front-bearing fraction, the occurrence rate is about 470 mini-waves per day over the Sun. The front energies were not measured.

\begin{figure}[t]
\centering
\includegraphics[width=0.98\linewidth]{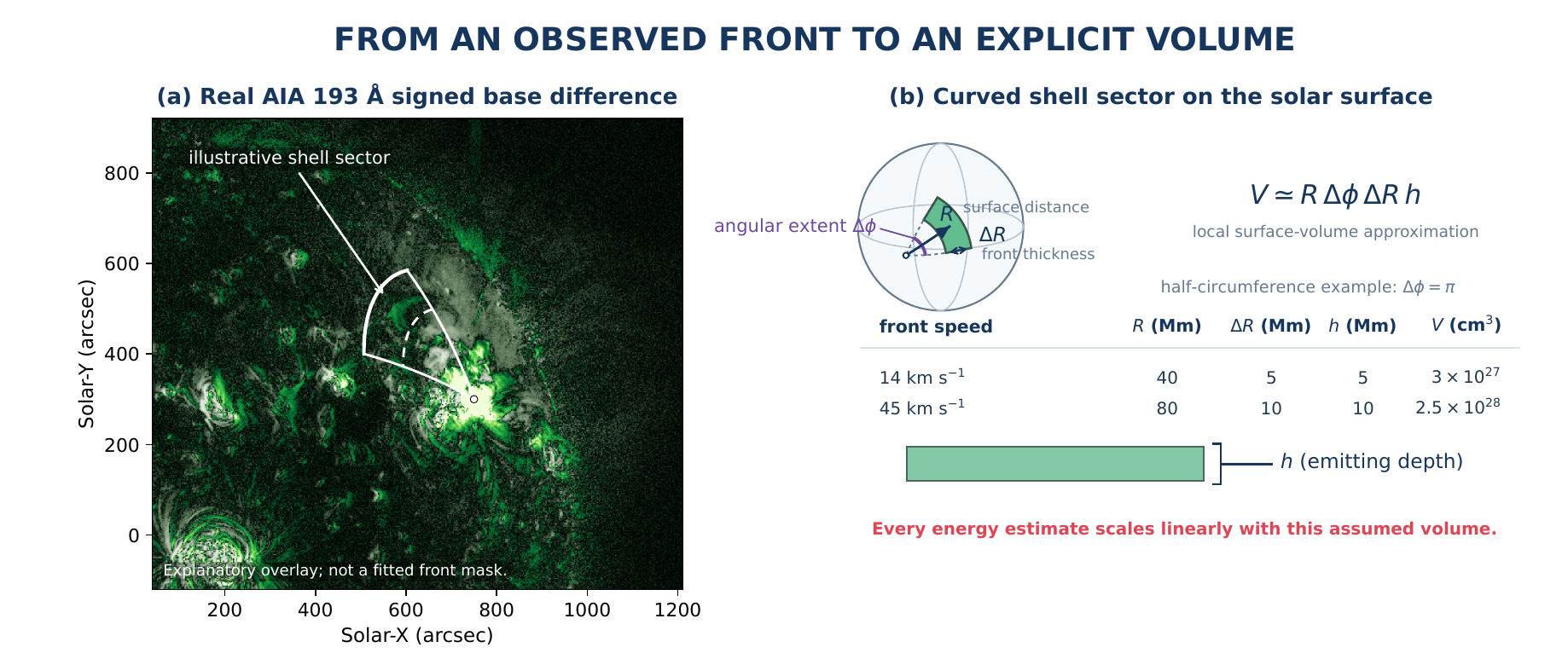}
\caption{Geometry of the illustrative mini-front estimate. (a) AIA 193~\AA{} signed base difference at 22:04~UT relative to 21:30~UT on 1 April 2017, with an explanatory shell sector centered on the apparent eruption site; its boundaries are not a fitted front mask. (b) Adopted volume $V\simeq R\,\Delta\phi\,\Delta R\,h$, where $\Delta\phi$ is the angular extent, $\Delta R$ the front thickness, and $h$ the emitting depth. The table gives the half-circumference examples used in Eq.~\eqref{eq:weakenergy}. The energy estimate changes linearly with this assumed volume. Green denotes the 193/195~\AA{} channel family only, not temperature.}
\label{fig:energygeometry}
\end{figure}

For a weak compressive disturbance, a simple kinetic estimate is
\begin{equation}
 E_{\rm kin}\simeq\frac{1}{2}\rho\,\delta v^2V,
 \qquad
 \delta v\simeq v_{\rm ph}\frac{\delta n}{n},
 \label{eq:weakenergy}
\end{equation}
where $v_{\rm ph}$ is the phase speed, not a bulk plasma speed. Taking $n_e=2\times10^8\,{\rm cm^{-3}}$, $\rho\simeq1.2m_p n_e$, a 5--10\% density compression, and illustrative shell-sector volumes of $3\times10^{27}$ and $2.5\times10^{28}\,{\rm cm^3}$ gives $3\times10^{21}$--$1.2\times10^{22}$ erg for the $14\,\kms$ event and $2.5\times10^{23}$--$1.0\times10^{24}$ erg for the $45\,\kms$ case. These are model-dependent order-of-magnitude estimates, not measured thermal energies and not a complete kinetic--thermal wave budget. The estimate is linear in density and volume and quadratic in both compression and phase speed, so its assumptions remain explicit and replaceable. Fig.~\ref{fig:energygeometry} makes the assumed shell-sector geometry explicit.

For comparison, the generic global-wave treatment of \textcite{PatsourakosVourlidas2012} gave about $1.8\times10^{29}$ erg when kinetic and changes in radiative/conductive energy fluxes were included, and explicitly described the result as a crude order-of-magnitude estimate. The independently diagnosed strong shock of \textcite{Long2015Energy} reached an estimated $2.8\times10^{31}$ erg. Purely geometrical reduction of the global-wave proxy to a mini-front shell gives $10^{26}$--a few $10^{27}$ erg, but this is a generous upper proxy rather than a measurement: the old mini-wave events lack the DEM constraints needed to establish comparable density and temperature perturbations.

Compact-source measurements provide a separate comparison. The Solar Orbiter analysis of \textcite{Podladchikova2025Picoflares} gives background-subtracted thermal energies of approximately $10^{20}$--$10^{24}$ erg, with a statistically robust power-law tail beginning near $10^{22}$ erg under the adopted geometry. The apparent flattening below the fitted range is treated as likely incompleteness, not a demonstrated physical turnover. The overlap between the upper compact-source range and the illustrative mini-front range is therefore suggestive, but it does not equate two unlike energy components.

\subsection{A same-event partition pilot: what is measured and what is not}
\begin{figure}[t]
\centering
\includegraphics[width=0.90\linewidth]{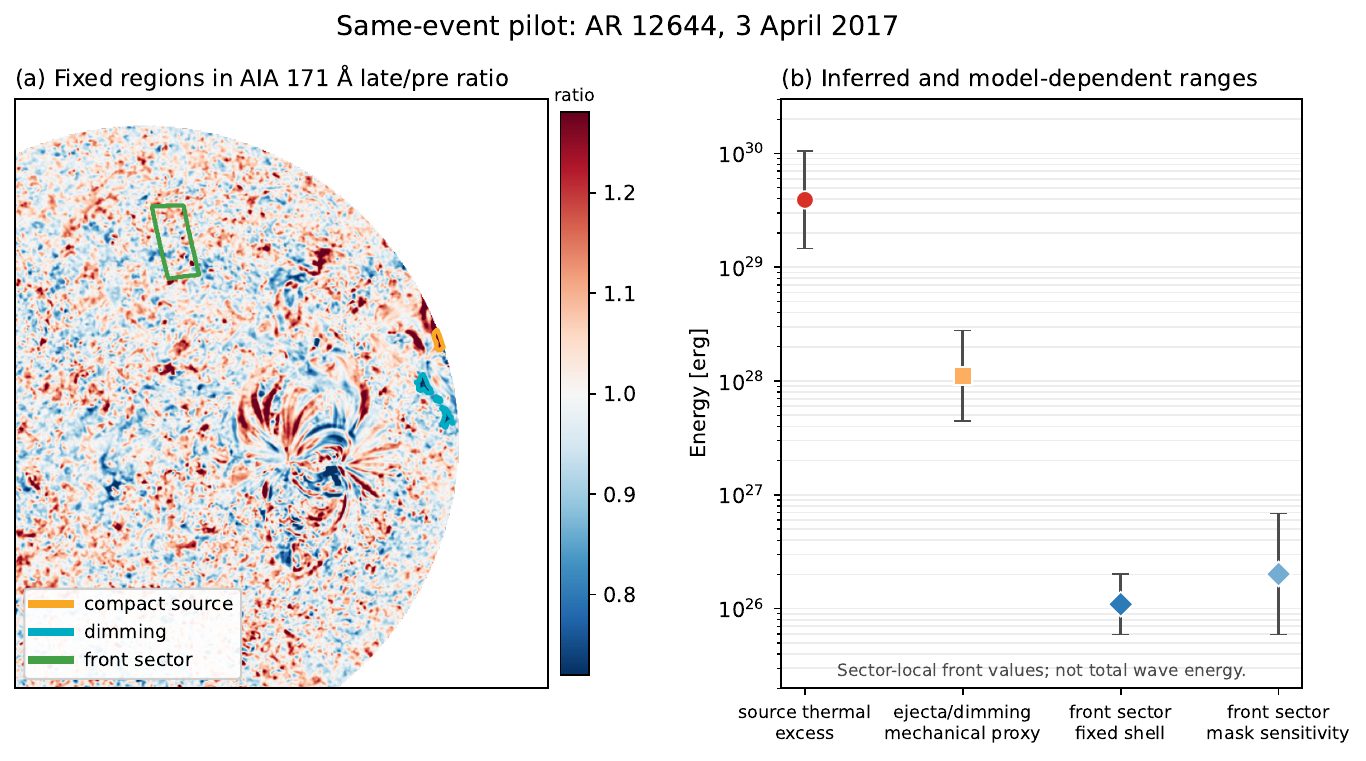}
\caption{Same-event partition pilot for AR~12644 on 3 April 2017. (a) Fixed source, dimming, and front-sector masks on the AIA 171~\AA{} 14:44/14:04 intensity ratio. The source estimate uses the 14:20 six-channel DEM; dimming and front use 14:44. The green front mask is reconstructed from the SWAP sector geometry analyzed by \textcite{OHara2019}; it is not an original O'Hara mask, a fitted crest contour, or a circular full-front extrapolation. Its weak contrast is consistent with a few-percent enhancement comparable to control-region variability. (b) Energy brackets from the same regions. Source thermal excess is DEM-inferred and requires a volume and filling factor; dimming/ejecta is a mechanical proxy; front values are weak-compression kinetic estimates for one sector. The brightness-selected subset is only an upper sensitivity case. These unlike quantities do not form a closed energy budget. The diverging palette in panel (a) represents decrease versus enhancement, not the 171/174~\AA{} channel-family color.}
\label{fig:partitionpilot}
\end{figure}

A same-event pilot can test this measurement on one larger eruption if all components share one geometry and observational quantities remain separate from model-dependent estimates. For the 3 April 2017 AR~12644 event, \textcite{OHara2019} located the source near N15W90 and tracked the northeastward on-disk wave in two sectors chosen to minimize active-region and coronal-hole interference. The present pilot retains this local sector geometry rather than imposing circular symmetry or extrapolating one segment around the Sun. Six exposure-normalized AIA channels were sampled at 14:04~UT (pre-event), 14:20~UT (early compact heating), and 14:44~UT (dimming and remote front). A hot-channel source mask, a strong 171~\AA{} dimming mask, and a fixed front-sector mask reconstructed from that SWAP sector geometry were transferred through one WCS; none was repositioned independently by passband. Fig.~\ref{fig:partitionpilot} shows the masks and energy brackets.

The direct image observables are the exposure-normalized intensities within these fixed regions. WCS-projected areas and the AIA 171~\AA{} ridge speed are derived quantities; DEM and energy estimates require further inversion and geometry. Six-channel regularized DEM inversion gives a source excess column emission measure of $2.1\times10^{28}\,\mathrm{cm^{-5}}$ at a DEM-weighted excess temperature near 9.7~MK, a late/pre dimming emission-measure ratio of 0.92, and a front-sector emission-measure ratio of 1.034. The latter is a 3.4\% enhancement, corresponding to $\delta n/n\simeq1.7$\% under constant line-of-sight depth. Consistently, the sector has only weak contrast in the 171~\AA{} ratio map. A brightness-selected subset gives $\delta n/n\simeq4.6$\%, but only as an upper sensitivity test because selection on the final enhancement biases the amplitude upward. Neighboring controls also vary by a few percent, placing the full-sector compression near the temporal-background floor rather than establishing a high-significance density measurement.

Energy conversion introduces the geometric model. Unit filling factor, a foreshortening-corrected source area, and depth $\sqrt{A}$ give an early compact-source thermal excess of $1.5\times10^{29}$--$1.0\times10^{30}$ erg. With 30--100~Mm depths, the strong-dimming mask gives a mass deficit of $2.1\times10^{12}$--$1.1\times10^{13}$ g. Combining that mass with the published 190--$370\,\kms$ off-limb-feature range gives a kinetic-plus-gravitational proxy of $4.4\times10^{27}$--$2.8\times10^{28}$ erg; associating that feature speed with the dimming mass is itself an assumption. Eq.~\eqref{eq:weakenergy}, using 405--$484\,\kms$ and 30--100~Mm depth, gives $5.9\times10^{25}$--$2.0\times10^{26}$ erg for the kinetic component in the fixed front segment. The brightness-selected sensitivity reaches $6.9\times10^{26}$ erg, but neither value is a full-front energy.

Relative to the inferred source thermal excess, the mechanical proxy is 0.4--19\%, the fixed front segment 0.006--0.14\%, and the brightness-selected sensitivity at most 0.47\%. These are not measurements of $E_{\rm ejecta}/\Erelease$ or $\Efront/\Erelease$: source thermal energy is only one release channel, the strongest dimming gives a lower mass bound, and the front mask covers one sector. The pilot therefore demonstrates a common event-level protocol but does not yet provide a point for fitting Eq.~\eqref{eq:universalscaling}.

\begin{figure}[t]
\centering
\includegraphics[width=0.98\linewidth]{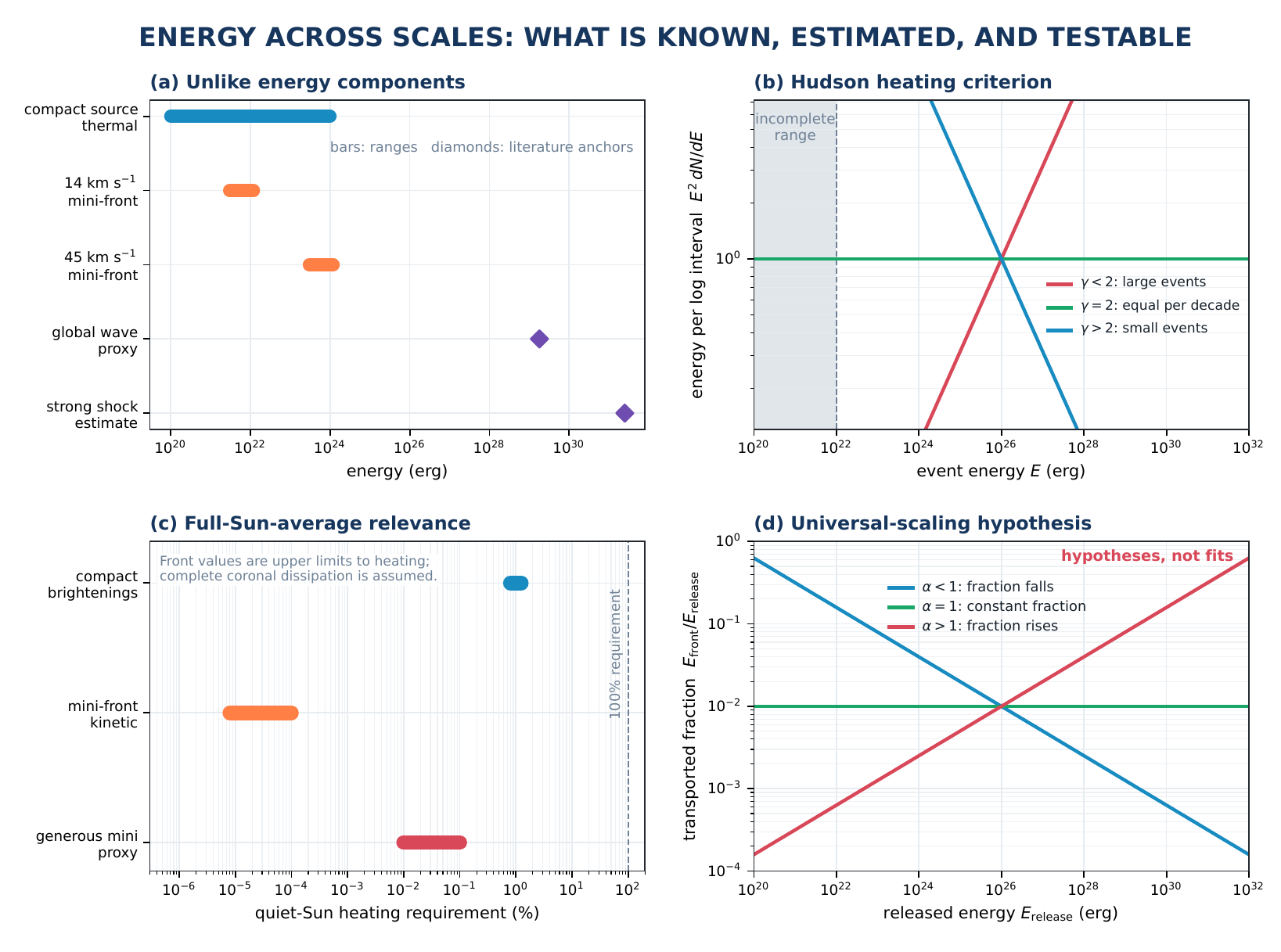}
\caption{Energetic synthesis and proposed test. (a) Source thermal energies, illustrative mini-front kinetic estimates, and global model estimates/proxies are shown separately because they are unlike quantities. (b) Energy per logarithmic interval for $dN/dE\propto E^{-\gamma}$; the shaded low-energy range is incomplete. (c) Full-Sun-average contributions relative to a $10^5\,{\rm erg\,cm^{-2}\,s^{-1}}$ quiet-Sun requirement; front values assume complete dissipation and are therefore upper limits to heating. (d) Three possible behaviors of Eq.~\eqref{eq:universalscaling}; the lines are hypotheses, not fits.}
\label{fig:energyscaling}
\end{figure}

Heating is best assessed as energy per logarithmic interval. If $dN/dE\propto E^{-\gamma}$, then $dQ/d\ln E\propto E^{2-\gamma}$ \parencite{Hudson1991}. Small events can dominate for $\gamma>2$ only if the power law continues below the completeness threshold to a physical lower cutoff; large events dominate for $\gamma<2$, and each decade contributes equally for $\gamma=2$. Fig.~\ref{fig:energyscaling} therefore marks the incomplete range rather than treating its flattening as physical.

The cross-scale question can be written as a candidate relation
\begin{equation}
 \Efront=C\Erelease^{\alpha},\qquad
 f_{\rm front}\equiv\frac{\Efront}{\Erelease}
 =C\Erelease^{\alpha-1}.
 \label{eq:universalscaling}
\end{equation}
Here $\alpha=1$ gives a constant transported fraction; $\alpha>1$ or $\alpha<1$ shifts that fraction toward larger or smaller eruptions. The present measurements do not justify a universal fit, but they define a falsifiable question: does $\Efront$ scale linearly with $\Erelease$, or does $\Efront/\Erelease$ change from compact quiet-Sun events to global waves and shocks? Combining the event distribution with Eq.~\eqref{eq:universalscaling} gives front-carried energy per logarithmic release-energy interval proportional to $E_{\rm release}^{\alpha+1-\gamma}$. With scale-independent dissipation, small events dominate only if $\gamma>\alpha+1$; $\gamma>2$ is recovered for $\alpha=1$. A valid fit requires paired same-event measurements with common definitions; the heterogeneous anchors in Fig.~\ref{fig:energyscaling} are not such a sample. Combined with front--CME separation and compression/Mach-number evolution, paired $\Erelease$ and $\Efront$ estimates can also test whether the front is energized mainly at launch or remains driven by CME expansion. EUV imaging, CME geometry, and radio diagnostics already show the value of such source--front tests \parencite{Ma2011Shock,Gopalswamy2012}. Neither $f_{\rm front}$ nor $\alpha$ alone identifies an MHD branch or a shock. For coronal heating, the key quantities are the fractions of released energy transported by the front and then dissipated in the corona.

Under the adopted assumptions, compact brightenings supply about 1\% of the quiet-Sun heating requirement. Mini-front energies of $10^{23}$--$10^{24}$ erg at 470 events per day give only $0.009$--$0.09\,{\rm erg\,cm^{-2}\,s^{-1}}$, or $8.9\times10^{-6}$--$8.9\times10^{-5}$\% of a $10^5\,{\rm erg\,cm^{-2}\,s^{-1}}$ requirement \parencite{WithbroeNoyes1977}. Even the generous $10^{26}$--$10^{27}$ erg proxy gives only $9$--$90\,{\rm erg\,cm^{-2}\,s^{-1}}$, or $0.009$--$0.09$\%. These are upper limits to heating because they assume that all transported front energy is dissipated in the quiet corona.

Uncertainty must precede any scaling fit. Because the weak-front estimate is quadratic in speed, a 20--30\% systematic speed error becomes a 44--69\% energy error; compression also enters quadratically, and volume uncertainty may dominate. Each event therefore requires one matched crest or leading edge, cadence-matched kinematics, DEM or spectroscopic compression, three-dimensional geometry, compact-source heating, and ejecta/dimming mass and kinetic energy.

Solar Orbiter can support this experiment: EUI resolves the source and faint front; SPICE constrains density, temperature, and Doppler perturbations; PHI supplies magnetic geometry; and Metis links low-coronal ejecta to CME evolution \parencite{Rochus2020,SPICE2020,Solanki2020,Antonucci2020}. Coordinated observations could turn Eq.~\eqref{eq:universalscaling} into an event-by-event test. This is an observing strategy, not a result claimed here.

\begin{figure}[!t]
\centering
\includegraphics[width=0.90\linewidth]{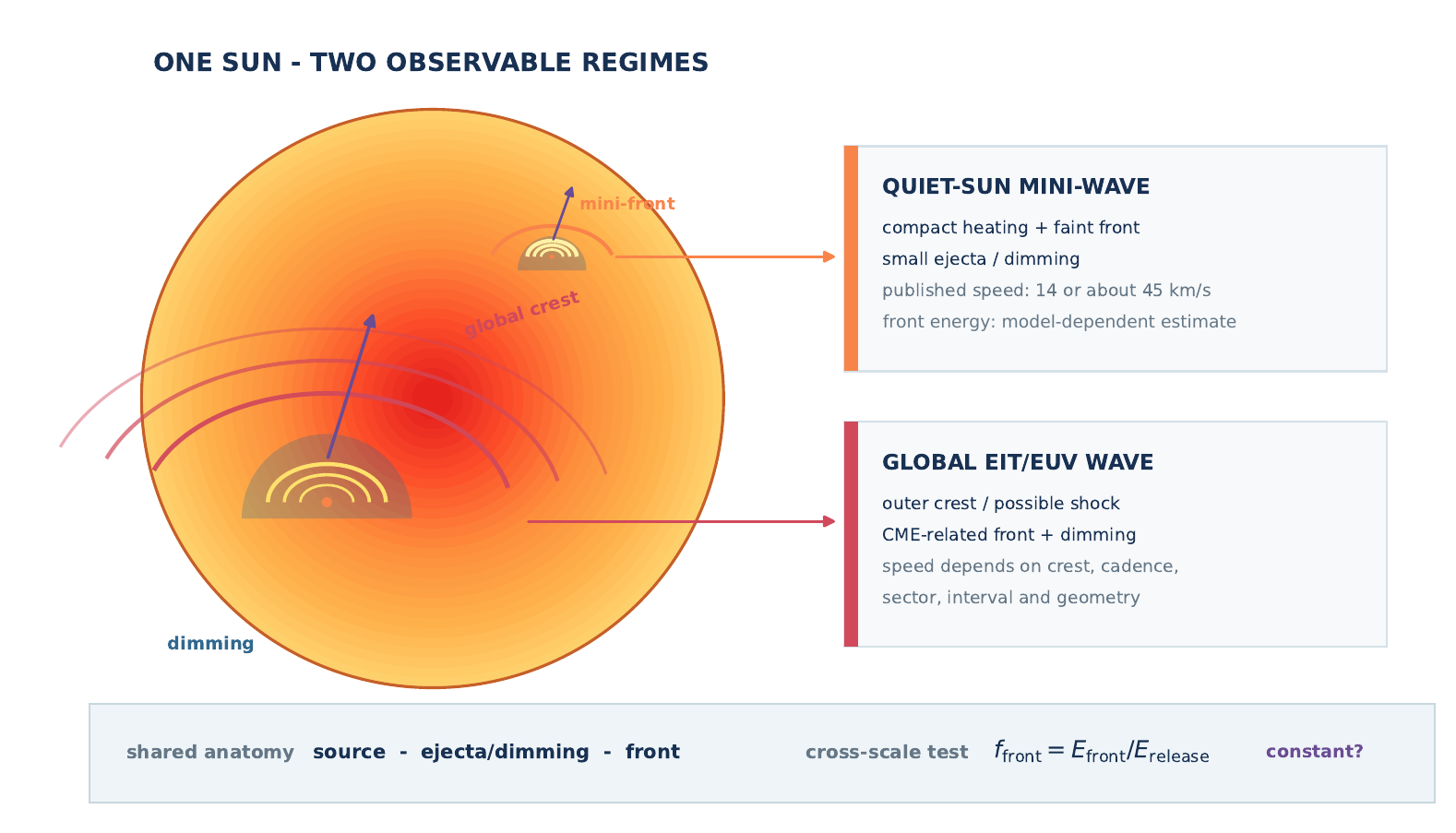}
\caption{Conceptual bridge between two observable regimes; the sketch is not to scale. Quiet-Sun mini-eruptions and global EIT/EUV events can share the source--ejecta/dimming--front anatomy without having the same MHD mode or energy partition. The speeds of 14 and about $45\,\kms$ are published mini-wave observables \parencite{Podladchikova2010,Innes2009}; their energy ranges are model-dependent weak-compression estimates introduced in this Perspective. A global outer wave or possible shock must likewise be separated from a CME-related front and dimming. The lower expression poses, but does not fit, the test in Eq.~\eqref{eq:universalscaling}: is $f_{\rm front}$ constant or scale-dependent?}
\label{fig:tworegimes}
\end{figure}

\section{Height matters, but a passband is not a height: connection to the solar wind}
Stereoscopy shows why projected distance is insufficient. \textcite{Tatiana2019} reconstructed two STEREO EUV-wave events and found front-emission heights of about 90--104~Mm in one event and an increase from about 54 to 93~Mm in another. Correcting for the evolving height changed the inferred speed by as much as 25\%. These heights correspond roughly to $1.08$--$1.15\,\Rs$: relevant to on-disk kinematics, but still in the low corona.

A 195~\AA{} image does not determine height: the passband weights temperature-sensitive emission along the line of sight rather than defining a geometrical shell. Height requires stereoscopy, limb constraints, tomography, spectroscopy, or another independent geometric diagnostic.

The same distinction limits connections to solar-wind acceleration. Localized momentum addition and rapid flux-tube divergence can produce multiple critical points and standing shocks; the sonic transition is set by the flow solution and tube geometry, not by an EUV wavelength \parencite{Habbal1985,Habbal1994}. EUV-wave heights can therefore locate low-coronal compression, heating, and wave-energy injection, but not the solar-wind sonic point. Testing a causal contribution to wind acceleration requires following that energy into the extended corona.

\section{A compact framework for future measurements}
Each reported front speed should identify the selected crest or leading edge and its time--distance ridge; passband, sector, fit interval, and native and effective cadence; WCS and projection treatment; relation to the CME flank and dimming; and the robustness of any signed azimuthal drift. A shock claim additionally requires an independent compression/Mach diagnostic. Fast/slow classification requires characteristic speeds and jump diagnostics, while quasi-parallel/quasi-perpendicular classification requires a three-dimensional front normal, an upstream magnetic-field estimate, and the resulting $\theta_{Bn}$, with model-dependent inputs labeled as such. Cadence, geometry, and feature definition should be matched before passband differences are interpreted physically.

An energy-partition record should also give $n_e$, $\delta n/n$, front volume and uncertainty, compact-source thermal energy, an ejecta/dimming mass proxy, and the basis for estimating total released energy. A phase speed without compression does not determine $\Efront$; $\Efront$ without a same-event denominator does not determine $f_{\rm front}$. These fields make Eq.~\eqref{eq:universalscaling} falsifiable while keeping measured quantities separate from model-dependent conversions.

The Nitta, Muhr, and Warmuth collections provide a strong starting point for this harmonization. A common observation operator can first test whether their kinematic groupings persist and then ask whether any surviving groups separate in compression, Mach number, $\theta_{Bn}$, or driver coupling. Solar Demon already demonstrates joint automated detection of flare, dimming, and EUV-wave candidates in SDO/AIA data \parencite{KraaikampVerbeeck2015}. Retaining masks, thresholds, and tracked-feature definitions would make those detections suitable for the same validation protocol.

The available systems form complementary analysis layers rather than a single performance ranking. NEMO and Solar Demon retain the eruption context by treating the propagating front together with dimming and, for Solar Demon, flare information. CorPITA and AWARE emphasize reproducible on-disk front detection and kinematics; AWARE uses RANSAC, a robust regression procedure that reduces the influence of outlying distance--time points. Wavetrack supplies multiscale feature segmentation and tracked morphology rather than an EUV-wave speed catalogue. SOLERwave combines single- and multi-sector measurements to construct a two-dimensional on-disk pattern-velocity field on an assumed solar-surface geometry. PyThea instead fits three-dimensional geometrical models to multi-view EUV and white-light images of CMEs and candidate shock envelopes. CASHeW moves the analysis mainly to near- and off-limb fronts and combines front tracking with differential-emission-measure and coronal-model information to estimate compression and shock-related quantities. The later SPREAdFAST study extends this approach outward by coupling a time-dependent front surface to coronal MHD, particle-acceleration, and SEP-transport modeling \parencite{Long2014CorPITA,Ireland2019AWARE,Stepanyuk2022Wavetrack,BaumgartnerSteinleitner2026SOLERwave,Kouloumvakos2022PyThea,Kozarev2017CASHeW,Kozarev2022SPREAdFAST}. These products go well beyond one scalar speed, but none alone provides direct measurements of every plasma variable or identifies a unique MHD shock family.

\clearpage
\begin{table}[p]
\centering
\caption{Complementary EUV-front samples and analysis frameworks. Numbers refer to the cited published products, not to one common benchmark. ``Model-assisted'' means that observational measurements are combined with a coronal model; it does not mean that the listed quantity is directly observed.}
\label{tab:frontframeworks}
\begingroup
\scriptsize
\renewcommand{\arraystretch}{1.08}
\resizebox{\textwidth}{!}{%
\begin{tabularx}{1.14\textwidth}{@{}>{\raggedright\arraybackslash}p{2.75cm}>{\raggedright\arraybackslash}p{3.25cm}Y Y@{}}
\toprule
\textbf{Study / product} & \textbf{Selection and geometry} & \textbf{Principal output} & \textbf{Verified relation or limitation} \\
\midrule
Warmuth--Mann \parencite{WarmuthMann2011} & Seventeen selected EUVI events and 61 kinematically usable EIT events from a 176-event EIT catalogue. & Initial speed and acceleration; three kinematic populations. & A physical kinematic classification, not an automated detection or shock-parameter catalogue. \\
Nitta et al. \parencite{Nitta2013} & 171 visually identified AIA 193~\AA{} LCPFs; 138 on-disk events analyzed in the paper. & Highest early directional speed and flare/CME/type-II associations. & The independently selected Nitta and Muhr tables contain nine identifiable common physical events, but only six have usable speeds in both studies; the later expanded 410-event Nitta online list supplied the input candidates for the CorPITA study. \\
Muhr et al. \parencite{Muhr2014} & 60 strong, clearly pronounced EUVI fronts; weak cases were excluded. & Perturbation-profile kinematics, amplitude, width, and eruption associations. & The six mutually speed-measurable events were used in the published comparison; 15 additional events overlap the Nitta 2014 analysis, giving the published 21-event speed comparison. \\
Solar Demon \parencite{KraaikampVerbeeck2015} & Automated SDO/AIA eruption monitoring rather than a fixed sample comparable to the three rows above. & Flare, dimming, and EUV-wave candidate detections within one system. & Its distinctive strength here is front--dimming context; it is not primarily a shock-parameter estimator. \\
CorPITA / Long et al. \parencite{Long2014CorPITA,Long2017Stat} & The expanded 410-event Nitta list was reprocessed; 362 on-disk events were analyzable and 164 were classified as waves. & Automated AIA 211~\AA{} sector profiles, front positions, velocities, accelerations, and event associations. & This is a direct algorithmic reanalysis of Nitta input candidates, not an independent event-discovery sample; the 164/362 result exposes sensitivity to the operational front definition. \\
AWARE \parencite{Ireland2019AWARE} & An automated method demonstrated on synthetic data and four AIA events. & Persistence maps and RANSAC-based distance, speed, and acceleration estimates. & A method and prototype, not a verified operational public catalogue and not an MHD shock classifier. \\
Wavetrack \parencite{Stepanyuk2022Wavetrack} & A multiscale image-processing and feature-tracking framework demonstrated on remotely sensed coronal structures; not a catalogue. & Wavelet-derived feature masks and evolving morphology. & Addresses feature definition and persistence, but does not infer a shock, Mach number, or MHD family. \\
SOLERwave \parencite{BaumgartnerSteinleitner2026SOLERwave} & An on-disk method demonstrated for the 6 September 2011 AIA wave; one event, not a catalogue. & Single- and multi-sector propagation estimates and a two-dimensional surface pattern-velocity field. & The reported $750$--$1500\,\kms$ range and $>40\%$ directional variation demonstrate within-event sector dependence; the vectors are pattern velocities on an assumed surface, not plasma velocities or MHD-mode labels. \\
PyThea \parencite{Kouloumvakos2022PyThea} & Interactive multi-view EUV and white-light reconstruction of selected CMEs and candidate shock envelopes. & GCS CME fits and ellipsoid/spheroid front geometry and kinematics. & Directly addresses projection and front-normal geometry; a fitted envelope alone does not establish a physical shock or its MHD family. \\
CASHeW \parencite{Kozarev2017CASHeW} & A framework for selected near- and off-limb coronal bright fronts rather than a general on-disk catalogue. & Front, peak, and back kinematics; evolving geometry; DEM/density compression; and model-assisted $M_A$ and $\theta_{Bn}$. & Moves from speed measurement toward physical shock characterization, but its inferred quantities remain conditional on the selected feature and coronal model. \\
SPREAdFAST \parencite{Kozarev2022SPREAdFAST} & SEP-conditioned selection beginning with 216 ERNE proton events; 62 cases retained measurable near/off-limb bright fronts. & Extension of the front surface through the corona with global-MHD sampling, particle-acceleration, and SEP-transport modeling. & A powerful Sun-to-1~AU extension, not a general EUV-wave population; its event selection and derived shock/particle quantities remain model dependent. \\
\bottomrule
\end{tabularx}%
}
\endgroup
\end{table}
\clearpage

The original papers explicitly establish two catalogue relationships: the Nitta--Muhr comparison of six mutually speed-measurable events within nine identifiable common physical events, and the direct CorPITA reprocessing of the expanded Nitta candidate list. A strict date-plus-flare-class cross-match of the published event tables for this revision additionally finds 20 shared eruptions between Nitta 2013 and SPREAdFAST, four between Muhr 2014 and SPREAdFAST, and three between Warmuth--Mann and SPREAdFAST. Two possible Nitta--SPREAdFAST matches, on 7 March and 9 April 2012, were not counted because the published metadata disagree. These are eruption-level matches, not proof that the analyses measured the same part of the front. A defensible front-level cross-match must also identify the source region, the same propagating structure, the viewing geometry, and the fitted ridge. Table~\ref{tab:frontframeworks} therefore compares what the products determine without claiming unverified one-to-one front equivalence.

For the same reason, the published speed distributions should not be compared as though they were measurements of one common coordinate. Warmuth--Mann emphasize initial speed and acceleration, Nitta retains the highest early on-disk directional speed, Muhr follows perturbation profiles over a longer evolution, and CorPITA fits automated on-disk sector profiles. SOLERwave instead returns a spatially resolved two-dimensional on-disk pattern-velocity field for one event. Its reported $750$--$1500\,\kms$ range and more than 40\% directional variation directly demonstrate that one scalar speed can depend on the sampled part of a front, but this within-event range is not a catalogue distribution. CASHeW and SPREAdFAST track the projected radial and lateral expansion of selected near- or off-limb fronts and use those tracks to constrain model-assisted geometry and shock parameters; PyThea supplies a separate geometrical reconstruction rather than a comparable speed catalogue. A difference among these outputs can therefore arise from selection, viewing geometry, propagation direction, evolutionary phase, or front definition before it implies a different physical population. Quantitative comparison requires a dedicated common-event reanalysis and is not attempted here.

For space-weather use, a large-scale EUV wave is strong evidence that a CME has occurred, even before coronagraph confirmation \parencite{Green2026}. Direction, rotation, and dimming topology may constrain the low-coronal CME footprint and possibly its trajectory or helicity-related evolution \parencite{Attrill2014}; their value for predicting geomagnetic impact or the magnetic-field orientation at Earth remains a testable statistical question for future catalogs.

\section{Conclusions}

Nearly three decades of EUV-wave observations do not require a single universal speed, because ``the speed'' is not a uniquely defined observable. The physical structure, MHD response, driver history, and observation operator are distinct parts of the inference; an EUV wavelength has neither an intrinsic Alfv\'en speed nor a geometrical height. Across the Nitta, Muhr, and Warmuth collections, cadence, sector, fitted interval, and feature selection change the reported observable. Warmuth's three populations, Nitta's absence of a comparable multi-class distribution, and Muhr's smooth transition within a selected strong-event sample therefore do not define competing MHD classifications. Direct Nitta--Muhr offsets and the repeated $834\,\kms$ value---numerically one SWAP radial ring per image interval---quantify the effect. Together, the three collections provide a useful basis for harmonized reanalysis: a kinematic grouping that survives common definitions of crest, sector, interval, projection, and uncertainty becomes a physical result worth explaining. Validated automated detections can supply the needed statistics if masks, thresholds, and feature definitions are retained; adding common-protocol compression and geometry would make the cross-scale energy and heating test feasible.

The modern progression is therefore not simply from a small catalogue to a larger one. Solar Demon preserves the coupled flare--front--dimming morphology; CorPITA and AWARE make on-disk detection and kinematics more reproducible; Wavetrack formalizes multiscale feature definition; SOLERwave resolves an on-disk two-dimensional pattern-velocity field; PyThea constrains multi-view three-dimensional geometry; and CASHeW and SPREAdFAST add near/off-limb thermodynamic context and model-assisted shock parameters. Their outputs answer different questions and should be connected through a common observational record rather than ranked by one detection rate. In particular, a fitted $M_A$ or $\theta_{Bn}$ is a powerful shock constraint, but it remains conditional on the selected front, density diagnostic, geometry, and coronal model.

Once the observation operator is controlled, classical theory provides a basis for specific, falsifiable tests rather than labels attached to velocity bins. Whitham's wavefront geometry \parencite[Sec.~7.9, pp.~259--262, Figs.~7.14--7.15]{Whitham1974}, the Kulikovskiy--Lyubimov discontinuity maps \parencite[Chap.~V, Secs.~1--2]{KulikovskiyLyubimov1965}, Korobeinikov's magnetized point-blast framework \parencite[Chap.~I, Secs.~3.3 and 4.1; Chap.~VII, Secs.~4--5]{Korobeinikov1973,Korobeinikov1991}, and the self-similar and weak-shock treatments applied by Grechnev and Afanasyev--Uralov \parencite{Grechnev2011,AfanasyevUralov2011} provide complementary analytical limits. The compression--Mach relation of \textcite{Vrsnak2002} shows how this works: under its stated low-$\beta$, perpendicular-shock assumptions, a density jump permits a quantitative shock test. None of these theories makes image-plane radial speed or azimuthal pattern drift an MHD-mode coordinate. Joint measurements of both, combined with compression, $\theta_{Bn}$, three-dimensional geometry, and front--CME separation, can nevertheless reject incompatible solutions. Two co-temporal, spatially distinct compressive fronts are therefore useful fast/slow-response candidates only after a CME-related front and observation-operator artifacts have been excluded; identifying a fast/slow shock pair additionally requires characteristic-speed and jump diagnostics for both fronts. Remote sensing can therefore move from visual resemblance to model discrimination.

The energy audit adds a cross-scale test. The mini-front kinetic energies near $10^{22}$--$10^{24}$ erg are heterogeneous, model-dependent weak-compression anchors, not a homogeneous sample for fitting a universal law, and their full-Sun-average contribution is far below the quiet-Sun heating requirement under the stated assumptions. The 3 April 2017 pilot nevertheless shows that compact-source heating, dimming depletion and a propagating-front segment can be measured within one WCS and DEM protocol. It also defines the current limit: the few-percent compression is comparable to control-region variability, while line-of-sight depth, full-front angular extent and total magnetic release remain unmeasured. The resulting brackets are therefore a controlled lower-bound experiment, not a closed energy partition.

The central experiment is thus neither a universal speed nor a power law already measured. It is whether $\Efront=C\Erelease^{\alpha}$ with $\alpha=1$, or whether $f_{\rm front}=\Efront/\Erelease$ changes systematically with scale. The exponent measures eruption-to-front coupling: $\alpha=1$ gives a constant transported fraction, whereas $\alpha>1$ or $\alpha<1$ shifts that fraction toward larger or smaller eruptions. Together with the event-energy distribution and the fraction dissipated in the corona, this determines which eruption scales can contribute materially to coronal heating. The main conclusion of this Perspective is therefore simple: standardize what the images measure, combine motion with compression, geometry, and driver evolution, and then let the data discriminate among modes, shocks, and energy-partition models. With that discipline, the diversity of measured speeds stops being a contradiction and becomes physical information. Fig.~\ref{fig:tworegimes} summarizes this cross-scale test.

The practical next step is not another speed-only catalogue. It is a common, openly documented event record that preserves source and dimming masks, the selected front or ridge, native cadence and passband, fit interval, projection, compression, three-dimensional geometry, model provenance, and uncertainties. Such a record would allow the same events to be reprocessed from on-disk EUV detection through off-limb and white-light shock characterization, while keeping measured quantities separate from model-dependent inference.

\section*{Acknowledgements}

The author is especially grateful to David Berghmans for introducing her to the EIT-wave problem and for initiating their early work on the quantitative characterization of wave and dimming observables---at that time a pioneering effort to move from visual descriptions toward measurable physical properties. She also thanks Ronald Van der Linden for recognizing the broader physical significance of this approach and for his longstanding scientific support. The author gratefully acknowledges stimulating scientific discussions with Leon Ofman, Valery M. Nakariakov, and Marco Velli on the interpretation of EIT/EUV disturbances, MHD modes, and the relation between propagating fronts and coronal structure. She also thanks Li Feng for a valuable discussion of Sedov--Taylor self-similar blast-wave solutions, which motivated the comparison with Korobeinikov's magnetized point-blast framework. Historical NEMO and conference material developed over many years provided motivation for revisiting the front--dimming problem with modern diagnostics.

\section*{Data availability}
This article combines a conceptual synthesis of published results with a controlled same-event AIA pilot. Its accompanying reproducibility archive is available at Zenodo, \url{https://doi.org/10.5281/zenodo.22288289}, and contains the final figure-generation, speed-audit, DEM and energy-estimator code; machine-readable masks, calculation inputs and outputs; a compact AIA temperature-response product; and a reduced public AIA Level-1.5 subset. Exact historical SWAP Level-1 files and original front masks are not included; reconstructed sector geometry is labeled as such rather than presented as an original mask. The AIA--SWAP speed comparison is therefore an observation-operator diagnostic, and the sector-local front energy is not presented as a total wave energy.

\section*{Software availability}
The public code repository is available at \url{https://github.com/epodlad/untangling-eit-waves}; its versioned v1.0.0 archive is included in the Zenodo record above. The package contains the reusable modules, scripts, manuscript source and figure assets needed to reproduce the calculations and figures in this review. It excludes local environments, calibration caches, temporary renders and development logs. Release manifests record filenames, sizes and SHA-256 checksums.

\section*{Competing interests}
The author declares no competing interests.

\section*{Author contributions}
O.P. conceived the synthesis, developed the physical framework, carried out the literature analysis, and wrote the manuscript.

\bibliographystyle{plainnat}
\scriptsize
\renewcommand{\baselinestretch}{0.92}\selectfont
\setlength{\bibsep}{0pt}
\bibliography{references}

@article{Thompson1998,
  author = {Thompson, B. J. and Plunkett, S. P. and Gurman, J. B. and Newmark, J. S. and St. Cyr, O. C. and Michels, D. J. and Delaboudiniere, J.-P.},
  title = {SOHO/EIT observations of an Earth-directed coronal mass ejection on 1997 May 12},
  journal = {Geophysical Research Letters},
  year = {1998},
  volume = {25},
  pages = {2465--2468},
  doi = {10.1029/98GL50429}
}

@article{Thompson1999,
  author = {Thompson, B. J. and Gurman, J. B. and Neupert, W. M. and Newmark, J. S. and Delaboudiniere, J.-P. and St. Cyr, O. C. and Stezelberger, S. and Dere, K. P. and Howard, R. A. and Michels, D. J.},
  title = {SOHO/EIT Observations of the 1997 April 7 Coronal Transient: Possible Evidence of Coronal Moreton Waves},
  journal = {Astrophysical Journal Letters},
  year = {1999},
  volume = {517},
  pages = {L151--L154},
  doi = {10.1086/312030}
}

@article{PodladchikovaBerghmans2005,
  author = {Podladchikova, O. and Berghmans, D.},
  title = {Automated Detection of EIT Waves and Dimmings},
  journal = {Solar Physics},
  year = {2005},
  volume = {228},
  pages = {265--284},
  doi = {10.1007/s11207-005-5373-z}
}

@article{Podladchikova2010,
  author = {Podladchikova, O. and Vourlidas, A. and Van der Linden, R. A. M. and Wuelser, J.-P. and Patsourakos, S.},
  title = {Extreme Ultraviolet Observations and Analysis of Micro-Eruptions and Their Associated Coronal Waves},
  journal = {Astrophysical Journal},
  year = {2010},
  volume = {709},
  pages = {369--376},
  doi = {10.1088/0004-637X/709/1/369}
}

@article{Podladchikova2012,
  author = {Podladchikova, O. and Vuiets, A. and Leontiev, P. and Van der Linden, R. A. M.},
  title = {Recent Developments of NEMO: Detection of EUV Wave Characteristics},
  journal = {Solar Physics},
  year = {2012},
  volume = {276},
  pages = {479--490},
  doi = {10.1007/s11207-011-9894-3}
}

@article{Tatiana2019,
  author = {Podladchikova, T. and Veronig, A. M. and Dissauer, K. and Temmer, M. and Podladchikova, O.},
  title = {3D Reconstructions of EUV Wavefront Heights and Their Influence on Wave Kinematics},
  journal = {Astrophysical Journal},
  year = {2019},
  volume = {877},
  eid = {68},
  doi = {10.3847/1538-4357/ab1b3a}
}

@article{NakariakovVerwichte2005,
  author = {Nakariakov, V. M. and Verwichte, E.},
  title = {Coronal Waves and Oscillations},
  journal = {Living Reviews in Solar Physics},
  year = {2005},
  volume = {2},
  eid = {3},
  doi = {10.12942/lrsp-2005-3}
}

@article{WillsDavey2007,
  author = {Wills-Davey, M. J. and DeForest, C. E. and Stenflo, J. O.},
  title = {Are ``EIT Waves'' Fast-Mode MHD Waves?},
  journal = {Astrophysical Journal},
  year = {2007},
  volume = {664},
  pages = {556--562},
  doi = {10.1086/519013}
}

@article{WarmuthMann2011,
  author = {Warmuth, A. and Mann, G.},
  title = {Kinematical Evidence for Physically Different Classes of Large-scale Coronal EUV Waves},
  journal = {Astronomy \& Astrophysics},
  year = {2011},
  volume = {532},
  eid = {A151},
  doi = {10.1051/0004-6361/201116685}
}

@article{Warmuth2015,
  author = {Warmuth, A.},
  title = {Large-scale Waves and Shocks in the Solar Corona},
  journal = {Living Reviews in Solar Physics},
  year = {2015},
  volume = {12},
  eid = {3},
  doi = {10.1007/lrsp-2015-3}
}

@article{Chen2002,
  author = {Chen, P. F. and Wu, S. T. and Shibata, K. and Fang, C.},
  title = {Evidence of EIT and Moreton Waves in Numerical Simulations},
  journal = {Astrophysical Journal Letters},
  year = {2002},
  volume = {572},
  pages = {L99--L102},
  doi = {10.1086/341486}
}

@article{Chen2005,
  author = {Chen, P. F. and Fang, C. and Shibata, K.},
  title = {A Full View of EIT Waves},
  journal = {Astrophysical Journal},
  year = {2005},
  volume = {622},
  pages = {1202--1210}
}

@article{ZhukovAuchere2004,
  author = {Zhukov, A. N. and Auchere, F.},
  title = {On the Nature of EIT Waves, EUV Dimmings and Their Link to CMEs},
  journal = {Astronomy \& Astrophysics},
  year = {2004},
  volume = {427},
  pages = {705--716},
  doi = {10.1051/0004-6361:20040351}
}

@article{Attrill2007,
  author = {Attrill, G. D. R. and Harra, L. K. and van Driel-Gesztelyi, L. and Demoulin, P.},
  title = {Coronal ``Wave'': Magnetic Footprint of a Coronal Mass Ejection?},
  journal = {Astrophysical Journal Letters},
  year = {2007},
  volume = {656},
  pages = {L101--L104},
  doi = {10.1086/512854}
}

@article{Nitta2013,
  author = {Nitta, N. V. and Schrijver, C. J. and Title, A. M. and Liu, W.},
  title = {Large-scale Coronal Propagating Fronts in Solar Eruptions as Observed by the Atmospheric Imaging Assembly on Board the Solar Dynamics Observatory: An Ensemble Study},
  journal = {Astrophysical Journal},
  year = {2013},
  volume = {776},
  eid = {58},
  doi = {10.1088/0004-637X/776/1/58}
}

@article{LiuOfman2014,
  author = {Liu, W. and Ofman, L.},
  title = {Advances in Observing Various Coronal EUV Waves in the SDO Era and Their Seismological Applications},
  journal = {Solar Physics},
  year = {2014},
  volume = {289},
  pages = {3233--3277},
  doi = {10.1007/s11207-014-0528-4}
}

@article{Long2008,
  author = {Long, D. M. and Gallagher, P. T. and McAteer, R. T. J. and Bloomfield, D. S.},
  title = {The Kinematics of a Globally Propagating Disturbance in the Solar Corona},
  journal = {Astrophysical Journal Letters},
  year = {2008},
  volume = {680},
  pages = {L81--L84},
  doi = {10.1086/589742}
}

@article{Long2017,
  author = {Long, D. M. and Bloomfield, D. S. and Chen, P. F. and Downs, C. and Gallagher, P. T. and Kwon, R.-Y. and Vanninathan, K. and Veronig, A. M. and Vourlidas, A. and Vrsnak, B. and Warmuth, A. and Zimovets, I. V. and others},
  title = {Understanding the Physical Nature of Coronal ``EIT Waves''},
  journal = {Solar Physics},
  year = {2017},
  volume = {292},
  eid = {7},
  doi = {10.1007/s11207-016-1030-y}
}

@article{OHara2019,
  author = {O'Hara, Jennifer P. and Mierla, Marilena and Podladchikova, Olena and D'Huys, Elke and West, Matthew J. and Talpeanu, Dana-Camelia and Rodriguez, Luciano and Zhukov, Andrei N.},
  title = {Exceptional Extended-field Observations of {EUV} Waves and Their Relation to the Accompanying Eruptions},
  journal = {The Astrophysical Journal},
  year = {2019},
  volume = {883},
  number = {1},
  eid = {59},
  doi = {10.3847/1538-4357/ab3b08}
}

@article{Grechnev2011,
  author = {Grechnev, V. V. and Uralov, A. M. and Chertok, I. M. and Kuzmenko, I. V. and Afanasyev, A. N. and Meshalkina, N. S. and Kalashnikov, S. S. and Kubo, Y.},
  title = {Coronal Shock Waves, {EUV} Waves, and Their Relation to {CMEs}. I. Reconciliation of ``{EIT Waves}'', Type {II} Radio Bursts, and Leading Edges of {CMEs}},
  journal = {Solar Physics},
  year = {2011},
  volume = {273},
  number = {2},
  pages = {433--460},
  doi = {10.1007/s11207-011-9780-z}
}

@article{AfanasyevUralov2011,
  author = {Afanasyev, A. N. and Uralov, A. M.},
  title = {Coronal Shock Waves, {EUV} Waves, and Their Relation to {CMEs}. II. Modeling {MHD} Shock Wave Propagation Along the Solar Surface, Using Nonlinear Geometrical Acoustics},
  journal = {Solar Physics},
  year = {2011},
  volume = {273},
  number = {2},
  pages = {479--491},
  doi = {10.1007/s11207-011-9730-9}
}

@book{Whitham1974,
  author = {Whitham, G. B.},
  title = {Linear and Nonlinear Waves},
  series = {Pure and Applied Mathematics},
  publisher = {Wiley-Interscience},
  address = {New York},
  year = {1974},
  isbn = {0-471-94090-9},
  doi = {10.1002/9781118032954}
}

@book{KulikovskiyLyubimov1965,
  author = {Kulikovskiy, A. G. and Lyubimov, G. A.},
  title = {Magnetohydrodynamics},
  publisher = {Addison-Wesley},
  address = {Reading, MA},
  year = {1965},
  pages = {216},
  isbn = {0201039508}
}

@book{Sedov1959,
  author = {Sedov, L. I.},
  title = {Similarity and Dimensional Methods in Mechanics},
  publisher = {Academic Press},
  year = {1959}
}

@article{KorobeinikovKarlikov1960,
  author = {Korobeinikov, V. P. and Karlikov, V. P.},
  title = {On the Interaction of Strong Explosion Waves with an Electromagnetic Field},
  journal = {Doklady Akademii Nauk SSSR},
  year = {1960},
  volume = {133},
  number = {4},
  pages = {764--767}
}

@article{Korobeinikov1973,
  author = {Korobeinikov, V. P.},
  title = {Problems in the Theory of Point Explosions in Gases},
  journal = {Proceedings of the Steklov Institute of Mathematics},
  year = {1973},
  volume = {119},
  pages = {1--311},
  url = {https://www.mathnet.ru/eng/tm3106}
}

@book{Korobeinikov1991,
  author = {Korobeinikov, V. P.},
  title = {Problems of Point Blast Theory},
  publisher = {American Institute of Physics},
  address = {New York},
  series = {AIP Translation Series},
  year = {1991},
  isbn = {0-88318-674-8}
}

@incollection{Petschek1964,
  author = {Petschek, H. E.},
  title = {Magnetic Field Annihilation},
  booktitle = {The Physics of Solar Flares},
  editor = {Hess, W. N.},
  series = {NASA Special Publication},
  volume = {50},
  pages = {425--439},
  publisher = {National Aeronautics and Space Administration},
  address = {Washington, DC},
  year = {1964}
}

@article{KraaikampVerbeeck2015,
  author = {Kraaikamp, E. and Verbeeck, C.},
  title = {Solar Demon---an Approach to Detecting Flares, Dimmings, and {EUV} Waves on {SDO/AIA} Images},
  journal = {Journal of Space Weather and Space Climate},
  volume = {5},
  pages = {A18},
  year = {2015},
  doi = {10.1051/swsc/2015019}
}

@inproceedings{PodladchikovaBerghmans2005Energetic,
  author = {Podladchikova, O. and Berghmans, D.},
  title = {Energetic Dynamics of EIT Wave Structure Analyzed by EIT Wave Detector},
  booktitle = {Connecting Sun and Heliosphere: Proceedings of Solar Wind 11 / SOHO 16},
  series = {ESA Special Publication},
  volume = {592},
  pages = {751--754},
  year = {2005}
}

@article{Vanninathan2018,
  author = {Vanninathan, K. and Veronig, A. M. and Dissauer, K. and Temmer, M.},
  title = {Plasma Diagnostics of Coronal Dimming Events},
  journal = {Astrophysical Journal},
  year = {2018},
  volume = {857},
  eid = {62},
  doi = {10.3847/1538-4357/aab09a}
}

@article{Dissauer2018,
  author = {Dissauer, K. and Veronig, A. M. and Temmer, M. and Podladchikova, T. and Vanninathan, K.},
  title = {Statistics of Coronal Dimmings Associated with Coronal Mass Ejections. I. Characteristic Dimming Properties and Flare Association},
  journal = {Astrophysical Journal},
  year = {2018},
  volume = {863},
  eid = {169},
  doi = {10.3847/1538-4357/aad3c6}
}

@article{Liu2010,
  author = {Liu, W. and Nitta, N. V. and Schrijver, C. J. and Title, A. M. and Tarbell, T. D.},
  title = {First SDO AIA Observations of a Global Coronal EUV ``Wave'': Multiple Components and ``Ripples''},
  journal = {Astrophysical Journal Letters},
  year = {2010},
  volume = {723},
  pages = {L53--L59},
  doi = {10.1088/2041-8205/723/1/L53}
}

@article{Liu2012,
  author = {Liu, W. and Ofman, L. and Nitta, N. V. and Aschwanden, M. J. and Schrijver, C. J. and Title, A. M. and Tarbell, T. D.},
  title = {Quasi-periodic Fast-mode Wave Trains Within a Global EUV Wave and Sequential Transverse Oscillations Detected by SDO/AIA},
  journal = {Astrophysical Journal},
  year = {2012},
  volume = {753},
  eid = {52},
  doi = {10.1088/0004-637X/753/1/52}
}

@article{Muhr2014,
  author = {Muhr, N. and Veronig, A. M. and Kienreich, I. W. and Vr{\v{s}}nak, B. and Temmer, M. and Bein, B. M.},
  title = {Statistical Analysis of Large-scale EUV Waves Observed by STEREO/EUVI},
  journal = {Solar Physics},
  year = {2014},
  volume = {289},
  pages = {4563--4588},
  doi = {10.1007/s11207-014-0594-7}
}

@article{Long2017Stat,
  author = {Long, D. M. and Murphy, P. and Graham, G. and Carley, E. P. and P{\'e}rez-Su{\'a}rez, D.},
  title = {A Statistical Analysis of the Solar Phenomena Associated with Global EUV Waves},
  journal = {Solar Physics},
  year = {2017},
  volume = {292},
  eid = {185},
  doi = {10.1007/s11207-017-1206-0}
}

@article{Veronig2010,
  author = {Veronig, A. M. and Muhr, N. and Kienreich, I. W. and Temmer, M. and Vr{\v{s}}nak, B.},
  title = {First Observations of a Dome-shaped Large-scale Coronal Extreme-ultraviolet Wave},
  journal = {Astrophysical Journal Letters},
  year = {2010},
  volume = {716},
  pages = {L57--L62},
  doi = {10.1088/2041-8205/716/1/L57}
}

@article{Cohen2009,
  author = {Cohen, O. and Attrill, G. D. R. and Manchester, W. B. IV and Wills-Davey, M. J.},
  title = {Numerical Simulation of an EUV Coronal Wave Based on the 2009 February 13 CME Event Observed by STEREO},
  journal = {Astrophysical Journal},
  year = {2009},
  volume = {705},
  pages = {587--602},
  doi = {10.1088/0004-637X/705/1/587}
}

@article{Cohen2010,
  author = {Cohen, O. and Attrill, G. D. R. and Schwadron, N. A. and Crooker, N. U. and Owens, M. J. and Downs, C. and Gombosi, T. I.},
  title = {Numerical Simulation of the 12 May 1997 CME Event: The Role of Magnetic Reconnection},
  journal = {Journal of Geophysical Research: Space Physics},
  year = {2010},
  volume = {115},
  eid = {A10104},
  doi = {10.1029/2010JA015464}
}

@article{Lulic2013,
  author = {Luli{\'c}, S. and Vr{\v{s}}nak, B. and {\v{Z}}ic, T. and Kienreich, I. W. and Muhr, N. and Temmer, M. and Veronig, A. M.},
  title = {Formation of Coronal Shock Waves},
  journal = {Solar Physics},
  year = {2013},
  volume = {286},
  pages = {509--528},
  doi = {10.1007/s11207-013-0287-7}
}

@article{Innes2009,
  author = {Innes, D. E. and Genetelli, A. and Attie, R. and Potts, H. E.},
  title = {Quiet Sun mini-CMEs activated by supergranular flows},
  journal = {Astronomy \& Astrophysics},
  volume = {495},
  pages = {319--323},
  year = {2009},
  doi = {10.1051/0004-6361:200811011}
}

@article{Ma2011Shock,
  author = {Ma, Suli and Raymond, John C. and Golub, Leon and Lin, Jun and Chen, Huadong and Grigis, Paolo and Testa, Paola and Long, David},
  title = {Observations and Interpretation of a Low Coronal Shock Wave Observed in the {EUV} by the {SDO/AIA}},
  journal = {Astrophysical Journal},
  volume = {738},
  number = {2},
  pages = {160},
  year = {2011},
  doi = {10.1088/0004-637X/738/2/160}
}

@article{Gopalswamy2012,
  author = {Gopalswamy, N. and Nitta, N. and Akiyama, S. and M{\"a}kel{\"a}, P. and Yashiro, S.},
  title = {Coronal Magnetic Field Measurement from {EUV} Images Made by the Solar Dynamics Observatory},
  journal = {Astrophysical Journal},
  volume = {744},
  number = {1},
  pages = {72},
  year = {2012},
  doi = {10.1088/0004-637X/744/1/72}
}

@article{Long2015Energy,
  author = {Long, D. M. and Baker, D. and Williams, D. R. and Carley, E. P. and Gallagher, P. T. and Zucca, P.},
  title = {The Energetics of a Global Shock Wave in the Low Solar Corona},
  journal = {The Astrophysical Journal},
  volume = {799},
  pages = {224},
  year = {2015},
  doi = {10.1088/0004-637X/799/2/224}
}

@article{Vanninathan2015,
  author = {Vanninathan, K. and Veronig, A. M. and Dissauer, K. and Madjarska, M. S. and Hannah, I. G. and Kontar, E. P.},
  title = {Coronal response to an EUV wave from DEM analysis},
  journal = {The Astrophysical Journal},
  volume = {812},
  pages = {173},
  year = {2015},
  doi = {10.1088/0004-637X/812/2/173}
}

@article{Delaboudiniere1995,
  author = {Delaboudiniere, J.-P. and Artzner, G. E. and Brunaud, J. and others},
  title = {EIT: Extreme-Ultraviolet Imaging Telescope for the SOHO Mission},
  journal = {Solar Physics},
  year = {1995},
  volume = {162},
  pages = {291--312},
  doi = {10.1007/BF00733432}
}

@article{PatsourakosVourlidas2012,
  author = {Patsourakos, S. and Vourlidas, A.},
  title = {On the Nature and Genesis of EUV Waves: A Synthesis of Observations from SOHO, STEREO, SDO, and Hinode},
  journal = {Solar Physics},
  year = {2012},
  volume = {281},
  pages = {187--222},
  doi = {10.1007/s11207-012-9988-6}
}

@article{Attrill2014,
  author = {Attrill, G. D. R. and Long, D. M. and Green, L. M. and Harra, L. K. and van Driel-Gesztelyi, L.},
  title = {Extreme-Ultraviolet Observations of Global Coronal Wave Rotation},
  journal = {Astrophysical Journal},
  year = {2014},
  volume = {796},
  eid = {55},
  doi = {10.1088/0004-637X/796/1/55}
}

@article{Vrsnak2002,
  author = {Vrsnak, B. and Magdalenic, J. and Aurass, H. and Mann, G.},
  title = {Band-splitting of Coronal and Interplanetary Type II Bursts. II. Coronal Magnetic Field and Alfven Velocity},
  journal = {Astronomy \& Astrophysics},
  year = {2002},
  volume = {396},
  pages = {673--682},
  doi = {10.1051/0004-6361:20021413}
}

@article{Green2026,
  author = {Green, L. M. and James, A. W. and Ngampoopun, N. and others},
  title = {Identifying and Predicting Coronal Mass Ejection Occurrence: Observational Checklists for Space Weather Forecasters},
  journal = {Space Weather},
  year = {2026},
  volume = {24},
  eid = {e2025SW004700},
  doi = {10.1029/2025SW004700}
}

@article{Podladchikova2025Picoflares,
  author = {Podladchikova, O. and Warmuth, A. and Harra, L. and others},
  title = {Picoflares in the Quiet Solar Corona: Solar Orbiter Observations Halfway to the Sun},
  journal = {arXiv e-prints},
  year = {2025},
  eid = {arXiv:2510.10340},
  doi = {10.48550/arXiv.2510.10340}
}

@article{Habbal1985,
  author = {Habbal, S. R.},
  title = {The Formation of a Standing Shock in a Polytropic Solar Wind Model within 1--10 Solar Radii},
  journal = {Journal of Geophysical Research},
  year = {1985},
  volume = {90},
  pages = {199--204},
  doi = {10.1029/JA090iA01p00199}
}

@article{Habbal1994,
  author = {Habbal, S. R. and Hu, Y. Q. and Esser, R.},
  title = {Standing Shocks in a Two-fluid Solar Wind},
  journal = {Journal of Geophysical Research},
  year = {1994},
  volume = {99},
  pages = {8465--8478},
  doi = {10.1029/94JA00349}
}

@article{WithbroeNoyes1977,
  author = {Withbroe, G. L. and Noyes, R. W.},
  title = {Mass and Energy Flow in the Solar Chromosphere and Corona},
  journal = {Annual Review of Astronomy and Astrophysics},
  year = {1977},
  volume = {15},
  pages = {363--387},
  doi = {10.1146/annurev.aa.15.090177.002051}
}

@article{Hudson1991,
  author = {Hudson, H. S.},
  title = {Solar Flares, Microflares, Nanoflares, and Coronal Heating},
  journal = {Solar Physics},
  year = {1991},
  volume = {133},
  number = {2},
  pages = {357--369},
  doi = {10.1007/BF00149894}
}

@article{Rochus2020,
  author = {Rochus, P. and Auch\`ere, F. and Berghmans, D. and others},
  title = {The Solar Orbiter EUI Instrument: The Extreme Ultraviolet Imager},
  journal = {Astronomy \& Astrophysics},
  year = {2020},
  volume = {642},
  eid = {A8},
  doi = {10.1051/0004-6361/201936663}
}

@article{SPICE2020,
  author = {{SPICE Consortium} and Anderson, M. and Appourchaux, T. and Auch\`ere, F. and others},
  title = {The Solar Orbiter SPICE Instrument: An Extreme UV Imaging Spectrometer},
  journal = {Astronomy \& Astrophysics},
  year = {2020},
  volume = {642},
  eid = {A14},
  doi = {10.1051/0004-6361/201935574}
}

@article{Solanki2020,
  author = {Solanki, S. K. and del Toro Iniesta, J. C. and Woch, J. and others},
  title = {The Polarimetric and Helioseismic Imager on Solar Orbiter},
  journal = {Astronomy \& Astrophysics},
  year = {2020},
  volume = {642},
  eid = {A11},
  doi = {10.1051/0004-6361/201935325}
}

@article{Antonucci2020,
  author = {Antonucci, E. and Romoli, M. and Andretta, V. and others},
  title = {Metis: The Solar Orbiter Visible Light and Ultraviolet Coronal Imager},
  journal = {Astronomy \& Astrophysics},
  year = {2020},
  volume = {642},
  eid = {A10},
  doi = {10.1051/0004-6361/201935338}
}

@article{Long2014CorPITA,
  author = {Long, D. M. and Bloomfield, D. S. and Gallagher, P. T. and P{\'e}rez-Su{\'a}rez, D.},
  title = {{CorPITA}: An Automated Algorithm for the Identification and Analysis of Coronal ``{EIT Waves}''},
  journal = {Solar Physics},
  year = {2014},
  volume = {289},
  number = {9},
  pages = {3279--3295},
  doi = {10.1007/s11207-014-0527-5}
}

@article{Ireland2019AWARE,
  author = {Ireland, J. and Inglis, A. R. and Shih, A. Y. and Christe, S. and Mumford, S. J. and Hayes, L. A. and Thompson, B. J. and Hughitt, V. K.},
  title = {{AWARE}: An Algorithm for the Automated Characterization of {EUV} Waves in the Solar Atmosphere},
  journal = {Solar Physics},
  year = {2019},
  volume = {294},
  number = {11},
  eid = {158},
  doi = {10.1007/s11207-019-1505-8}
}

@article{Kozarev2017CASHeW,
  author = {Kozarev, K. A. and Davey, A. and Kendrick, A. and Hammer, M. and Keith, C.},
  title = {The Coronal Analysis of {SHocks} and Waves ({CASHeW}) Framework},
  journal = {Journal of Space Weather and Space Climate},
  year = {2017},
  volume = {7},
  eid = {A32},
  doi = {10.1051/swsc/2017028}
}

@article{Kozarev2022SPREAdFAST,
  author = {Kozarev, K. and Nedal, M. and Miteva, R. and Dechev, M. and Zucca, P.},
  title = {A Multi-Event Study of Early-Stage {SEP} Acceleration by {CME}-Driven Shocks---Sun to 1~{AU}},
  journal = {Frontiers in Astronomy and Space Sciences},
  year = {2022},
  volume = {9},
  eid = {801429},
  doi = {10.3389/fspas.2022.801429}
}

@article{Stepanyuk2022Wavetrack,
  author = {Stepanyuk, Oleg and Kozarev, Kamen and Nedal, Mohamed},
  title = {Multi-scale Image Preprocessing and Feature Tracking for Remote {CME} Characterization},
  journal = {Journal of Space Weather and Space Climate},
  year = {2022},
  volume = {12},
  pages = {20},
  doi = {10.1051/swsc/2022020}
}

@article{Kouloumvakos2022PyThea,
  author = {Kouloumvakos, Athanasios and Rodr\'iguez-Garc\'ia, Laura and Gieseler, Jan and Price, Daniel J. and Vourlidas, Angelos and Vainio, Rami},
  title = {{PyThea}: An Open-Source Software Package to Perform 3D Reconstruction of Coronal Mass Ejections and Shock Waves},
  journal = {Frontiers in Astronomy and Space Sciences},
  year = {2022},
  volume = {9},
  pages = {974137},
  doi = {10.3389/fspas.2022.974137}
}

@article{BaumgartnerSteinleitner2026SOLERwave,
  author = {Baumgartner-Steinleitner, Markus and Veronig, Astrid M. and Dissauer, Karin and Pomoell, Jens},
  title = {Investigation of the Two-Dimensional Velocity Field of the Large-Scale Coronal Wave from September 6, 2011 Using the {SOLERwave} Tool},
  journal = {Solar Physics},
  year = {2026},
  volume = {301},
  pages = {102},
  doi = {10.1007/s11207-026-02690-6}
}
\end{document}